\documentclass[12pt]{article}
\usepackage{latexsym}
\usepackage{epsfig,amssymb,euscript,slashed}
\usepackage[linktocpage]{hyperref}
\usepackage{amsmath}
\usepackage[noadjust]{cite}
\usepackage{mathtools}
\usepackage{physics}
\usepackage{extarrows}
\usepackage{xcolor}
\usepackage{float}
\usepackage{graphicx}
\usepackage{subcaption}
\usepackage{tcolorbox}
\usepackage{array,calc,epsfig}
\usepackage{bbm}
\usepackage{fancybox}

\usepackage{tikz}
\usepackage{adjustbox}
\usetikzlibrary{decorations.markings}
\tikzset{->-/.style={decoration={
			markings,
			mark=at position #1 with {\arrow{stealth}}},postaction={decorate}}}
\tikzset{photon/.style={decorate,decoration={snake}}}
\usepackage{tikz-feynman}

\newcommand{\hb}{\bar{h}}
\newcommand{\pd}{\partial}
\newcommand{\pdb}{\bar{\partial}}
\newcommand{\mb}{\bar{m}}
\newcommand{\zb}{\bar{z}}
\newcommand{\jb}{\bar{j}}
\newcommand{\Db}{\widetilde{D}}

\newcommand{\pp}{p_{+}}
\newcommand{\pn}{p_{-}}
\newcommand{\OL}{\mathcal{O}_L}
\newcommand{\OLb}{\bar{\mathcal{O}}_L}
\newcommand{\OH}{\mathcal{O}_H}

\newcommand{\Ofer}{\mathcal{O}^{\mathrm{fer}}}
\newcommand{\Obos}{\mathcal{O}^{\mathrm{bos}}}
\newcommand{\cIt}{\tilde{I}}
\newcommand{\C}[1]{C^{\,2}_{\!(#1)}}
\newcommand{\G}[1]{\Gamma_{\!(#1)}}
\newcommand{\D}[1]{\delta_{(#1)}}

\numberwithin{equation}{section} 
\usepackage[]{graphicx}
\usepackage{color}

\def\({\left(}
\def\){\right)}
\def\[{\left[}
\def\]{\right]}

\def\pd{{\partial}}

\def\One{{\hbox{ 1\kern-.8mm l}}}

\def\barray{\begin{array}}
\def\earray{\end{array}}
\def\be{\begin{equation}}
\def\ee{\end{equation}}
\def\bea{\begin{eqnarray}}
\def\eea{\end{eqnarray}}
\def\bal{\begin{align}}
\def\eal{\end{align}}

\def\mb{\bar{m}}

\numberwithin{equation}{section} 
\allowdisplaybreaks  

\makeatletter
\g@addto@macro\bfseries{\boldmath}
\makeatother

\definecolor{cardinal}{rgb}{0.6,0,0}
\definecolor{darkgreen}{rgb}{0,0.4,0}
\definecolor{golden}{rgb}{0.92, 0.7, 0}
\definecolor{midnight}{rgb}{0, 0, 0.5}
\definecolor{darkblue}{rgb}{0, 0, 0.7}

\def\cC{{\cal C}}

\def\cI{{\cal I}}

\def\cM{{\cal M}}
\def\cN{{\cal N}}

\def\cO{{\cal O}}

\def\bp{{\mathbf{p}}}

\def\zb{{\bar{z}}}
\def\zbar{{\bar{z}}}
\def\hb{{\bar{h}}}
\newcommand{\plog}[2]{\mathrm{Li}_{#1} \left( #2 \right) } 
\def\pdp{\partial_+}

\usepackage{epsf}
\usepackage{cite}
\usepackage{fancyhdr}
\usepackage{bm}
\usepackage{enumerate} 

\usepackage{empheq} \empheqset{box=\fbox}

\numberwithin{equation}{section}  
\allowdisplaybreaks

\begin{document}
\font\cmss=cmss10 \font\cmsss=cmss10 at 7pt

\begin{flushright}{  
\scriptsize QMUL-PH-21-10}
\end{flushright}
\hfill
\vspace{18pt}
\begin{center}
{\Large 
\textbf{The Regge limit of AdS$_3$ holographic correlators with\\\vspace{6pt} heavy states: towards the black hole regime
}}
\end{center}

\vspace{8pt}
\begin{center}
{\textsl{Nejc \v{C}eplak$^{\,a}$ and Marcel R. R. Hughes$^{\,b}$}}

\vspace{1cm}

\textit{\small ${}^a$ Institut de Physique Th\'eorique, 
	Universit\'e Paris Saclay,
	 CEA, CNRS, \\ 	Orme des Merisiers, Gif sur Yvette, 91191 CEDEX, France  \\}   \vspace{6pt}

\textit{\small ${}^b$ Centre for Research in String Theory, School of Physics and Astronomy\\
Queen Mary University of London,
Mile End Road, London, E1 4NS,
United Kingdom}\\
\vspace{6pt}

\end{center}

\vspace{12pt}

\begin{center}
\textbf{Abstract}
\end{center}

\vspace{4pt} {\small
\noindent

We examine the Regge limit of holographic 4-point correlation functions in AdS$_3\times S^3$ involving two heavy and two light operators. 
In this kinematic regime such correlators can be reconstructed from the bulk phase shift accumulated by the light probe as it traverses the geometry dual to the heavy operator. 
We work perturbatively -- but to arbitrary orders -- in the ratio of the heavy operator's conformal dimension to the dual CFT${}_2$'s central charge, thus going beyond the low order results of \cite{Kulaxizi:2018dxo} and \cite{Giusto:2020mup}.
In doing so, we derive all-order relations between the bulk phase shift and the Regge limit OPE data of a class of heavy-light multi-trace operators exchanged in the cross-channel.
Furthermore, we analyse two examples for which the relevant 4-point correlators are known explicitly to all orders: firstly the case of heavy operators dual to AdS${}_3$ conical defect geometries and secondly the case of non-trivial smooth geometries representing microstates of the two-charge D1-D5 black hole.

}

\vspace{1cm}

\thispagestyle{empty}

\vfill
\vskip 5.mm
\hrule width 5.cm
\vskip 2.mm
{
\noindent  {\scriptsize e-mails:  {\tt nejc.ceplak@ipht.fr, m.r.r.hughes@qmul.ac.uk} }
}

\setcounter{footnote}{0}
\setcounter{page}{0}

\newpage

\tableofcontents

\section{Introduction}
\label{sec:intro}

The scattering of two objects is arguably one of the simplest physical experiments that can be thought up. 
Despite apparent conceptual simplicity, the details of such a process contain a wealth of information about the nature of interactions between the scattering bodies and their internal structure. 
In order to isolate particular behaviours, it is common to resort to certain approximations or kinematical regimes to simplify the analysis. 
An example of this is the limit of high energy scattering at fixed momentum transfer -- the eikonal limit -- in which the contributions of individual Feynman diagrams can be approximately resummed into a single phase containing all details of the interaction in this regime \cite{Levy:1969cr}. 
Considerations of this regime have been particularly fruitful for theories that include gravity, since in this limit the gravitational interactions dominate \cite{Amati:1987wq,tHooft:1987vrq}.

The eikonal approximation can also be studied in the framework of the AdS/CFT duality \cite{Maldacena:1997re}, where 
high-energy scattering in the bulk can be related to the Regge limit of the corresponding CFT correlation functions \cite{Cornalba:2006xk,Cornalba:2006xm,Cornalba:2007zb,Cornalba:2007fs}.
In this work, we are interested in a class of scattering processes in which a light probe scatters from a fixed heavy target -- this we take to be a non-empty semiclassical geometry formed from the backreaction of the heavy object.
In the Regge regime the light probe is highly relativistic and its trajectory can be approximated by a null geodesic that begins and ends on the asymptotic AdS boundary of the geometry.
As the probe traverses the spacetime it accumulates a phase; the details of which contains rich information about the interaction between the probe and the heavy object. 
This phase shift can also be used to analyse the Regge limit of a CFT 4-point function containing two heavy and two light operators (HHLL), as demonstrated in \cite{Kulaxizi:2018dxo}.
In that work, this setup was used to extract the eikonal phase for AdS black holes and conical defects, which in the case of two-dimensional CFTs was then matched to the behaviour of the HHLL Virasoro vacuum block \cite{Fitzpatrick:2014vua,Fitzpatrick:2015zha} in the Regge limit, emphasising the dominance of the stress tensor sector in this kinematical regime. 
The bootstrap relations in the Regge limit and their implications for bulk physics were considered variously in \cite{Cornalba:2009ax,Costa:2012cb,Kulaxizi:2017ixa,Li:2017lmh,Karlsson:2019dbd,Karlsson:2019qfi,Li:2020dqm}, while stringy corrections were considered in~\cite{Cornalba:2008qf,Brower:2007xg,Meltzer:2019pyl,Antunes:2020pof}.
 
In \cite{Giusto:2020mup}, this Regge limit analysis was applied to correlators involving pure heavy operators dual to smooth horizonless geometries with the same conserved charges and asymptotic behaviour as the Strominger-Vafa D1-D5-P black hole \cite{Strominger:1997eq}.
In the framework of the fuzzball proposal \cite{Mathur:2005zp, Mathur:2008nj}, such geometries -- dual to pure states of the CFT -- are considered to be at least a subset of the aforementioned black hole's microstates.
Due to the lack of a closed form for the exact expressions of holographic 4-point correlation functions involving these heavy operators, the analysis of \cite{Giusto:2020mup} was limited to first order in a perturbative expansion in the ratio between the reduced conformal weight of the heavy state $h^{[0]}_H$ and the central charge $c$ of the CFT \cite{Kulaxizi:2018dxo}
\begin{align} \label{eq:introperturbRegime}
	\mu \sim \frac{h^{[0]}_H}{c} \quad\text{with}\quad h^{[0]}_H\sim O(c) \quad\text{as}\quad c\to\infty \,.
\end{align}
In the supergravity regime, HHLL correlators can be computed by extracting the 2-point function of the light operators in the geometry dual to the heavy operator \cite{Galliani:2016cai, Galliani:2017jlg}. While such an approach avoids the use of Witten diagrams, analytic results are often limited to only linear perturbations away from pure AdS \cite{Bombini:2017sge, Bombini:2019vnc}.\footnote{The case of linear perturbations gives rise to LLLL correlators in AdS${}_3$ which are, however, under good control~\cite{Rastelli:2019gtj}.} In the bulk, the deviation from global AdS is parametrised by $\mu$, which in principle can have an arbitrary fixed value, however, here we take it to be parametrically small and treat it as an expansion parameter. The scaling of the conformal dimension with the central charge is necessary in order to ensure that the heavy operator corresponds to a classical geometry.
For the relatively simple, though still quite non-trivial microstate geometry referred to as the (1,0,0) superstrata, some HHLL correlators can be evaluated exactly in $\mu$ in terms of a double Fourier series \cite{Bombini:2017sge}.
As is explored in greater detail in an upcoming publication \cite{CGHR}, expanding this result in $\mu$ we obtain at each order a closed form expression in terms of generalised Bloch-Wigner functions \cite{zagier1991polylogarithms} which here we use to extend the linear analysis of \cite{Giusto:2020mup} to higher orders.
For comparison, we review the analogous calculation for the case of the AdS spacetime with a conical defect of order $k$ in three bulk dimensions \cite{Kulaxizi:2018dxo,Karlsson:2019qfi}. When this $k$ is a positive integer, the geometry has a known dual pure heavy state \cite{Balasubramanian:2000rt,Maldacena:2000dr,Lunin:2001jy}. However, if $k$ is analytically continued to be real-valued this is no longer the case and the resulting geometry is instead thought to be dual to a mixed state in the CFT. 
One aim in the present paper is to examine whether it is possible to use the Regge limit analysis at higher orders in $\mu$ to distinguish between geometries dual to pure and mixed states; done by contrasting the $(1,0,0)$ microstate geometry and the analytically continued conical defect. 
Briefly summarising, while the Regge limit analysis is consistent at arbitrary orders in $\mu$ and so should be applicable even in the black hole regime, our analysis appears to suggest that this kinematical regime is unable to tell apart pure and mixed states.
Such a check is a prerequisite for the application of the Regge limit to analyse or predict correlators involving more complicated microstates, where explicit expressions for the correlation functions may not be known beyond the leading order in the parameter $\mu$. 
In fact, for the study of black hole physics the opposite limit -- that of large $\mu$ -- is of interest, since in that regime the 3-charge microstates closely resemble the classical black hole geometry and thus one expects higher orders in the series expansion in $\mu$ to contribute significantly.

A further goal is to understand the relationship of this Regge limit analysis at higher orders in the $\mu$ expansion (in the context of AdS/CFT) with work in flat space that is perturbative in $G_N$, see~\cite{Bern:2021dqo} and references therein. 
Throughout this paper, we limit the presentation of our analysis to order $\mu^3$, however, all results have been checked to higher orders.
We note that similar ideas have been explored in \cite{Karlsson:2019qfi}, but with an emphasis on the AdS-Schwarzschild geometries in arbitrary dimensions (or an AdS conical defect in the case of $D=3$), which in general are believed not to be dual to any particular pure state in the dual CFT. Wherever applicable, our results are in perfect agreement with  \cite{Karlsson:2019qfi}.

The structure of this paper is as follows.
In section~\ref{sec:back} we briefly review the geodesic approximation to the computation of the bulk phase shift, present the methods used to analyse the corresponding HHLL 4-point correlation functions in the Regge limit, and detail the relation between the bulk and CFT data with an emphasis on higher orders in the $\mu$ expansion.
These ideas are then applied to the examples of the AdS${}_3$ conical defect geometry in section~\ref{sec:conical} and the $(1,0,0)$ microstate in section~\ref{sec:twocharge}.
We conclude with a summary and discussion in section~\ref{sec:disc}. 
Some of the more technical details pertaining to the correlator in the two-charge geometry are collected in the appendix~\ref{app:corr}.

\section{Regge limit analysis}
\label{sec:back}

The physical process of interest here is the scattering of a highly energetic light particle from a heavy object in asymptotically AdS$_3$ spacetimes.
In the bulk, one can study such an experiment by approximating the path of the probe with a null geodesic in the spacetime obtained from the backreaction of the heavy object. 
On the other hand, this scattering can be equivalently described in the dual two-dimensional CFT: there one needs to analyse the Regge limit of a 4-point correlation function with two heavy and two light operators, dual to the classical spacetime and light probe respectively.

In this section we present the gravitational and field-theoretic set-up necessary to describe such a scattering process and present the proposal of \cite{Kulaxizi:2018dxo}, which relates the phase shift acquired by a null geodesic as it traverses an asymptotically AdS spacetime to the Regge limit of a 4-point correlation function in the dual CFT. 
By working perturbatively in the ratio of the conformal dimension of the heavy operator and the central charge of the CFT, we extend the results of \cite{Karlsson:2019qfi, Giusto:2020mup} to arbitrarily high orders and elucidate the connection between bulk and CFT data.

\subsection{The bulk phase shift}
\label{ssec:BulkPic}

Consider a three-dimensional spacetime that is asymptotically global AdS$_3$.
Let the metric be diagonal so that we can write the line element as%
\footnote{The signature of the metric is chosen to be mostly positive.}
\begin{align}
	\label{eq:lineelement3d}
	ds^2_{\!AdS} = g_{\mu\nu} \, dx^{\mu}dx^{\nu}= g_{tt}\, dt^2+ g_{yy} \,dy^2 + g_{rr}\, dr^2\,,
\end{align}
where $r$ denotes the radial coordinate, with the boundary of AdS located at $r \to \infty$. 
The time coordinate is denoted by $t$ and $y$ is a spatial coordinate which is periodic \mbox{$y \sim y + 2 \pi R_y\,$}, with $R_y$ denoting the radius of the circle at the boundary. 
These coordinates $(t,y)$ are also used to describe the cylinder on which we define the dual two-dimensional CFT. Assuming that the components of the metric \eqref{eq:lineelement3d} depend only on the radial coordinate, there exist two killing vectors $\pd_t$ and $\pd_y$ and their associated momenta
\begin{align} \label{eq:ptpyDef}
	p_t = g_{tt}\dot{t}\qquad\text{and}\qquad p_y = g_{yy}\dot{y} \,, 
\end{align}
which are conserved along the geodesic.\footnote{The dot represents a derivative with respect to the affine parameter of the null geodesic.}
It is convenient to define the ratio
\begin{align} \label{eq:betaDef}
	\beta \equiv \frac{p_y}{p_t} \,, 
\end{align}
which labels the null geodesics and is related to the flat space limit of the AdS impact parameter.
We note that for our choice of conventions $p_t\leq0$ while $p_y$ can take any sign and thus we have $\abs{\beta}\leq1$.

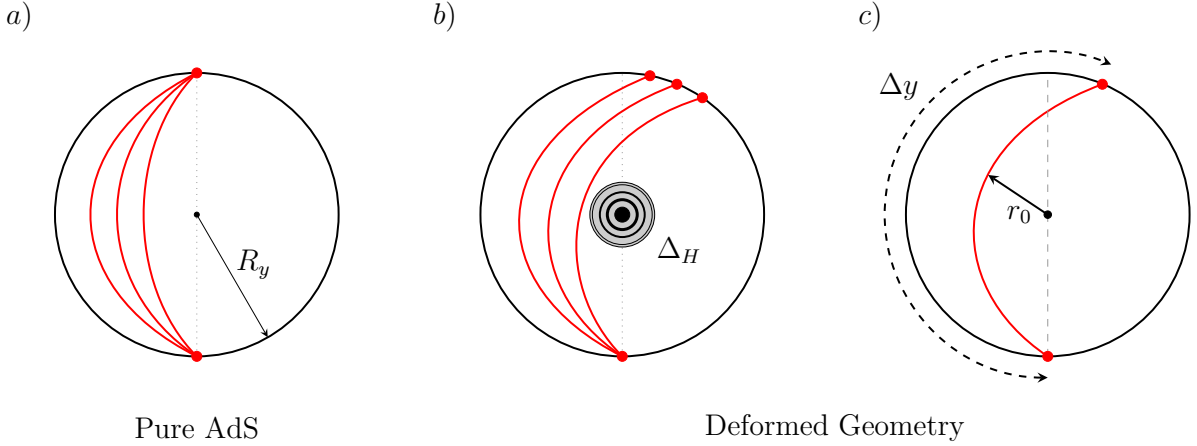
\begin{figure}[t]
	\begin{center}
		\begin{adjustbox}{max totalsize={\textwidth}{0.4\textheight},center}
			\begin{tikzpicture}
				\draw[black, thick] (0,0) circle  (2);
				\node[] at (0,-3) {Pure AdS};
				\draw[red, thick] (0,-2) .. controls (-1,-1) and (-1,1).. (0,2);
				\draw[red, thick] (0,-2)  .. controls (-1.5,-1) and (-1.5,1).. (0,2);
				\draw[red, thick] (0,-2)  .. controls (-2,-1) and (-2,1).. (0,2);
				\draw[gray!80, dotted] (0,-2) -- (0,2);
				\filldraw[red] (0,-2) circle (2pt);
				\filldraw[red] (0,2) circle (2pt);
				\node[] at (-2.5,2.8) {$a)$};
				\filldraw[black] (0,0) circle (1pt);
				\draw[black, ->, >=stealth] (0,0) -- (1, -1.73);
				\node[] at (0.8, -0.7) {$R_y$};
				\draw[black, thick] (6,0) circle  (2);
				\node[] at (9,-3) {Deformed Geometry};
				\draw[gray!80, dotted] (6,-2) -- (6,2);
				\filldraw[color=black, thin, fill = gray!40] (6,0) circle (13pt);
				\filldraw[black] (6,0) circle (3pt);
				\draw[black, very thick] (6,0) circle (6pt);
				\draw[black, thick] (6,0) circle (9pt);
				\draw[black] (6,0) circle (12pt);
				\node[] at (6.8,-0.5) {$\Delta_H$};
				\draw[red, thick] (6,-2) .. controls (4,-1) and (4,1).. (6+0.39,1.96);
				\draw[red, thick] (6,-2) .. controls (4.5,-1) and (4.5,1).. (6+0.77,1.84);
				\draw[red, thick] (6,-2) .. controls (5,-1) and (5,1).. (6+1.13,1.65);
				\filldraw[red] (6,-2) circle (2pt);
				\filldraw[red] (6+0.39,1.96) circle (2pt);
				\filldraw[red] (6+0.77,1.84) circle (2pt);
				\filldraw[red] (6+1.13,1.65) circle (2pt);
				\node[] at (6-2.5,2.8) {$b)$};
				\draw[black, thick] (12,0) circle  (2);
				\draw[red, thick] (12+0,-2) .. controls (12-1.5,-1) and (12-1.5,1).. (12+0.77,1.84);
				\filldraw[red] (12+0.77,1.84) circle (2pt);
				\draw[gray!80, dashed] (12+0,-2) -- (12+0,2);
				\filldraw[red] (12+0,-2) circle (2pt);
				\draw[ black, thick, ->, >={stealth}] (12+0,0) -- (12-1.5*0.56,1*0.56);
				\filldraw[black] (12+0,0) circle (1.5pt);
				\node[] at (12-0.4, 0) {$r_0$};
				\draw[black,thick,dashed, <->, >={stealth}] (12+0, -2.3) arc (270:90-22.91:2.3);
				\node[] at (12-2.1, 1.8) {$\Delta y$};
				\node[] at (12-2.5,2.8) {$c)$};
			\end{tikzpicture}
		\end{adjustbox}
	\caption{The projection of null geodesics in asymptotically AdS$_3$ spaces with time increasing out of the page and the asymptotic AdS boundary being the outer solid black circle. 
	In $a)$ we depict null geodesics in pure AdS which all converge to one point (with $\Delta y = \pi R_y$) given fixed initial boundary data.
	In $b)$ we show an asymptotically AdS spacetime with a schematic representation of the deformation of spacetime away from empty AdS. This causes the geodesics to be deflected and they re-emerge from the bulk at different points.  
	Figure $c)$ indicates the spatial shift $\Delta y$ acquired by a null geodesic as it crosses the bulk and the radial turning point $r_0$ as the point of closest approach to the origin $r=0$. 
	}
	\label{fig:shift}
	\end{center}
\end{figure}

To each null geodesic we can associate the observable
\begin{align} \label{eq:BulkPS}
	\delta(p_t, p_y) \equiv  -p_t \,\Delta t - p_y \,\Delta y= 2\,|p_t| \!\int_{r_0}^{\infty}\!dr\,\frac{\dot{t} + \beta\dot{y}}{\dot{r}}\,,
\end{align}
where $\Delta t$ and  $\Delta y$ are respectively the temporal and spatial  distances between the endpoints of the geodesic measured on the AdS boundary -- see figure~\ref{fig:shift}.
The integration is along the null geodesic with $r_0$ denoting its radial turning point -- the minimal radial distance of the path from the origin -- which can be obtained (if it exists) from the the minimum real solution to the null geodesic condition $g_{\mu\nu}\dot{x}^{\mu}\dot{x}^{\nu} = 0$, yielding for these geometries
\begin{align} \label{eq:rdoteq}
	\dot{r}^2 = p_t^{2}\, \frac{g_{yy} + \beta^2 g_{tt}}{ - g_{tt}g_{yy}g_{rr}} = 0 \,.
\end{align}
In the case of three-dimensional diagonal metrics we can write \eqref{eq:BulkPS} explicitly as
\begin{align} \label{eq:BulkPS2}
	\delta(p_t, p_y) = 2\,|p_t|\!\int_{r_0}^{\infty}\!dr\, \sqrt{g_{rr}} \,\sqrt{\frac{g_{yy} + \beta^2 g_{tt}}{- g_{tt}g_{yy}}\,} \ .
\end{align}
This observable is the eikonal phase shift of the light particle as it crosses the bulk geometry and contains information about the interaction between the probe and the heavy object, including possible capture effects \cite{Parnachev:2020zbr}. 
Equivalently, \eqref{eq:BulkPS2} encodes data about four-point correlation functions involving operators dual to the classical geometry and light probe. 
Note that in holographic calculations the quantity used is actually the difference $\delta - \delta_{\mathrm{AdS}}$, with $\delta_{\mathrm{AdS}}$ being the phase of a null geodesic in pure AdS$_3$.%
\footnote{For all such null geodesics in pure AdS$_3$ one finds $\Delta t = \Delta y = \pi R_y$ (see also figure~\ref{fig:shift}).}

\subsection{The Regge limit of 4-point CFT correlation functions}
\label{ssec:CFTPic}

Following \cite{Maldacena:1997re}, type IIB string theory on AdS${}_3 \times S^3 \times \cM$ (with $\cM$ denoting a compact four-dimensional manifold, taken to be either $T^4$ or $K3$) has an equivalent description in terms of a two-dimensional $\mathcal{N}=(4,4)$ supersymmetric CFT with $SU(2)_L\times SU(2)_R$ R-symmetry.
We are interested in the supergravity limit of this duality where, as the name suggests, the bulk is described in terms of classical supergravity and the dual theory is a strongly coupled CFT with a large central charge $c = 6N$.%
\footnote{The parameter $N=n_1n_5$ stems from the string theory construction of the so-called D1-D5 system, where a bound state of a large number $n_1$ of D1-branes and $n_5$ of D5-branes is considered. 
Thus from the relation $c= 6N$ we see that the large $c$ limit is equivalent to taking the large $N$ limit and we use the two limits interchangeably throughout the text.} 

In particular, this duality allows us to describe scattering processes in AdS$_3$ by using correlation functions in the CFT \cite{Giddings:1999jq, Giddings:1999qu, Polchinski:1999ry, Penedones:2010ue}. 
The elastic scattering of two particles translates into a 4-point correlation function containing two pairs of conjugate operators
\begin{align}
\langle \bar{\mathcal{O}}_1(z_1,\zb_1) {\mathcal{O}}_1(z_2,\zb_2) \mathcal{O}_2(z_3,\zb_3) \bar{\mathcal{O}}_2(z_4,\zb_4) \rangle\,.
\end{align}
Conformal invariance fixes the form of such a correlator up to an arbitrary function of two cross-ratios, which we take to be 
\begin{align} \label{eq:zzbDef}
	z = \frac{z_{14}z_{23}}{z_{13}z_{24}} \ \ , \qquad \zb = \frac{\zb_{14}\zb_{23}}{\zb_{13}\zb_{24}} \ .
\end{align}
Thus we can use conformal symmetry to fix the positions of three operator insertions as
\begin{align}
	z_1 \to 0\,, \qquad z_2 \rightarrow \infty\,, \qquad z_3 \rightarrow 1\,,
\end{align}
implying $z_4 \to z$, and so we will study instead the correlator
\begin{align}\label{eq:gaugefix}
	 \langle {\mathcal{O}}_1(\infty) \mathcal{O}_2(1)\bar{\mathcal{O}}_2(z,\zb) \bar{\mathcal{O}}_1(0)\rangle  = \lim_{\substack{z_2,\zb_2\to\infty\\z_1,\zb_1\to0\\z_3,\zb_3\to1}} z_2^{2h_1}\zb_2^{2\hb_1} \langle \bar{\mathcal{O}}_1(z_1,\zb_1) {\mathcal{O}}_1(z_2,\zb_2) \mathcal{O}_2(z_3,\zb_3) \bar{\mathcal{O}}_2(z_4,\zb_4) \rangle\,,
\end{align}
where $h_i$ and $\hb_i$ denote the holomorphic and anti-holomorphic conformal dimensions of the operator $\mathcal{O}_i$. 

The gravitational picture described in the previous subsection corresponds to a particular class of 4-point functions, dubbed heavy-heavy-light-light (HHLL), where two of the operators are taken to be `heavy' in the sense that in the large $N$ limit their dimensions scale with the central charge ($\Delta_H \sim O(c)$).
In contrast, the scaling dimensions of the remaining two operators are kept fixed in this limit ($\Delta_L \sim O(1)$) and thus we refer to them as `light'.
We denote such HHLL correlators of the conformally-fixed form \eqref{eq:gaugefix} as
\begin{align} \label{eq:HHLLCorr}
     \cC(z,\zb) \equiv \langle {\mathcal{O}}_H(\infty) \mathcal{O}_L(1)\bar{\mathcal{O}}_L(z,\zb) \bar{\mathcal{O}}_H(0)\rangle \ .
\end{align}
It is helpful for us to disentangle the Virasoro and $SU(2)$ current sectors using the Sugawara construction, from which results a reduced conformal dimension defined as (here given for the heavy operator)
\begin{align} \label{eq:ReducedDim}
    h_H^{[0]} \equiv h_H - \frac{j_H^2}{N} \ ,
\end{align}
where $j_H^2$ is the eigenvalue of the quadratic Casimir of $SU(2)_L$.%
\footnote{In \eqref{eq:ReducedDim} we used the notation for the conformal dimensions of heavy operators since under this procedure the reduced dimensions of light operators and the value of the central charge are unchanged at leading order in the large $N$ limit.}

	\begin{figure}[t]
		\begin{center}
		\begin{adjustbox}{max totalsize={\textwidth}{0.5\textheight},center}
			\begin{tikzpicture}
				\draw[black, thick] (0,0) circle  (2);
				\node[] at (2.12*0.85, 2.12*0.85)   {\large $\bar \cO_{H}$};
				\node[] at (-2.12*0.85, 2.12*0.85)   {\large $\cO_H$};
				\node[] at (-2.12*0.85, -2.12*0.85)   {\large $\cO_L$};
				\node[] at (2.12*0.85, -2.12*0.85)   {\large $ \bar \cO_L$};
				\draw[black, thick] (1.41,1.41) -- (0, 0.77); 
				\draw[black, thick] (-1.41,1.41) -- (0, 0.77); 
				\draw[black, thick] (-1.41,-1.41) -- (0,- 0.77); 
				\draw[black, thick] (1.41,-1.41) -- (0,- 0.77); 
				\draw[black, thick] (0,0.77) -- (0,- 0.77); 
				\node[] at (0.4,0.) { \large $ \cO'$};
				\node[ align=center] at (0,-3) {direct-channel};
				\node[] at (-3,2.5) {$a)$};
				\draw[black, thick] (6+2,0) circle  (2);
				\node[] at (2.12*0.85+6+2, 2.12*0.85)   {\large $\bar \cO_{H}$};
				\node[] at (-2.12*0.85+6+2, 2.12*0.85)   {\large $\cO_H$};
				\node[] at (-2.12*0.85+6+2, -2.12*0.85)   {\large $\cO_L$};
				\node[] at (2.12*0.85+6+2, -2.12*0.85)   {\large $\bar \cO_L$};
				\draw[black, thick] (1.41+6+2,1.41) -- (6+0.77+2,0); 
				\draw[black, thick] (-1.41+6+2,1.41) -- (6-0.77+2, 0); 
				\draw[black, thick] (-1.41+6+2,-1.41) -- (6-0.77+2,0); 
				\draw[black, thick] (1.41+6+2,-1.41) -- (6+0.77+2,0 ); 
				\draw[black, thick] (5+0.23+2,0) -- (7-0.23+2,0); 
				\node[] at (6+2,0.3) { \large $ \cO$};
				\node[align=center] at (6+2,-3) {cross-channel};
				\node[] at (-3+8,2.5) {$b)$};
			\end{tikzpicture}
		\end{adjustbox}
	\caption{A schematic expansion of the 4-point correlator into different channels. 
	In figure $a)$ we show the direct-channel ($z_1 \to z_2$ or equivalently $z \to 1$) where the two light operators in a HHLL correlator are brought together. 
	In figure $b)$ we depict the cross-channel expansion ($z_1 \to z_4$ or $z \to 0 $)  where the heavy and light operators are contracted. 
	While one can equivalently expand in either channel, the set of operators exchanged (denoted by $\cO'$ in the direct-channel and $\cO$ in the cross-channel) are generically different.
	}
	\label{fig:cft}
	\end{center}
	\end{figure}
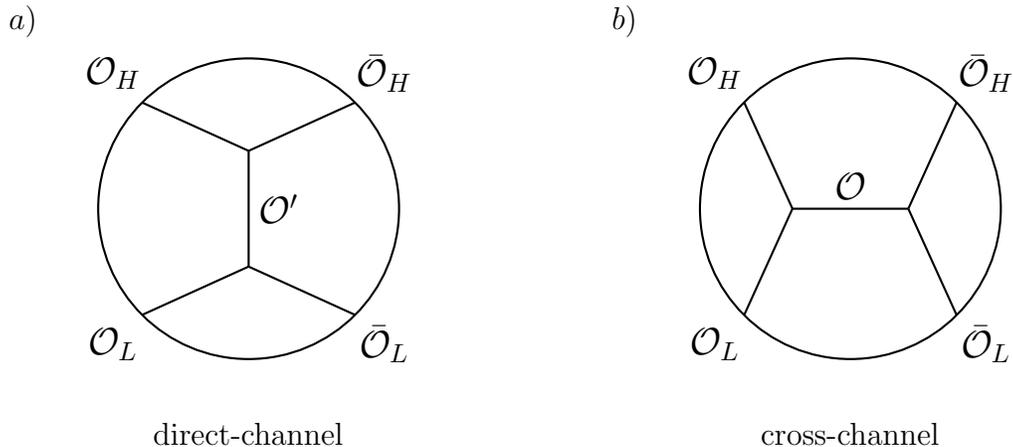

The correlator \eqref{eq:HHLLCorr} can be decomposed into different channels, for instance by summing over the exchanges of quasi-primary operators (see figure~\ref{fig:cft})
\begin{subequations}
	\label{eq:channeldec}
\begin{align} 
	 \cC(z,\zb) &= (1-z)^{-2h_L} (1-\bar{z})^{-2\bar{h}_L}\sum_{\mathcal{O}'} C_{HH\mathcal{O}'}C_{\bar{\mathcal{O}}'LL}\, g^{0,0}_{h,\hb}(1-z,1-\zb) \ ,
	\label{eq:DirectChannel}
	\\*
	 &= z^{-h^{[0]}_H-h_L}\zb^{-\hb^{[0]}_H-\hb_L}\sum_{\mathcal{O}} C_{H\!L\mathcal{O}}C_{\bar{\mathcal{O}}L\!H}\, g^{h_{H\!L},\bar{h}_{H\!L}}_{h,\hb}(z,\zb) \ ,\label{eq:CrossChannel}
\end{align}
\end{subequations}
where we define $h_{H\!L}\equiv h^{[0]}_H-h_L$ and use the global conformal blocks given by the usual hypergeometric form
\begin{align} 
	\label{eq:ConformalBlockDef}
	g^{a,\bar a}_{h,\hb}(z,\zb) = z^{h}\zb^{\hb} {}_2F_1\big(h-a,h-a;2h;z\big)\,{}_2F_1\big(\hb-\bar a,\hb-\bar a;2\hb;\zb\big)\,.
\end{align}

We refer to the OPE decomposition \eqref{eq:DirectChannel} where the two light operators are brought together by taking $z_1\to z_2$ (or equivalently $z\to1$) as the `direct-channel'. 
In this channel, both single-trace operators (the identity, stress tensor and current operators) and double-trace operators are exchanged.%
\footnote{As is common in the literature, we borrow nomenclature from the $\cN=4$ SYM theory in which single-trace operators are the single-particle states of the dual AdS${}_5$ theory and multi-trace operators denote composite objects made out of single-trace components, dual to multi-particle states.}
The double-trace operators exchanged in this channel are composed of two conjugate operators and schematically\footnote{This form of writing the double-trace operators is schematic, since at each level the derivatives are understood to act in such a way as to define a primary operator.} denoted by $\cO_{LL} \equiv \,:\!\bar{\mathcal{O}}_L\pd^m\pdb^{\mb}\mathcal{O}_L\!:$. We write these as LL double-trace operators since such operators with heavy constituents give subleading contributions to the correlator in the large $N$ limit.%

Likewise, we call the OPE limit where a heavy and a light operator are brought together \eqref{eq:CrossChannel} by taking $z_1\to z_4$ ($z\to0$) the `cross-channel'.
In this channel, no single-trace operators are exchanged and the leading contribution in $1/N$ comes from the exchange of a family of HL multi-trace\footnote{We use the terminology of `multi-trace' for these HL composite operators since in general the heavy operator itself can -- and for us will -- be a multi-trace operator.} operators which we denote by
\begin{align} \label{eq:OHL}
	\mathcal{O}_{H\!L} \equiv\,:\! \mathcal{O}_{\!H}\pd^m\pdb^{\mb}\mathcal{O}_{\!L}\!: \ .
\end{align}
The conformal dimensions of these operators are given by
\begin{align}\label{eq:DTConDim}
	h = h^{[0]}_H + h_L + m + \frac{1}{2}\Gamma_{m,\mb} \ ,\qquad \hb= \hb^{[0]}_H+\hb_L+\mb+\frac{1}{2}\Gamma_{m,\mb} \ ,
\end{align}
with $\Gamma_{m,\mb}$ denoting the anomalous dimensions, which can be thought of as the analogues of binding energies between the different constituents of the multi-trace operator.
With the conformal dimensions of the external operators being fixed, at leading order in $1/N$ the sum over operators exchanged in the cross-channel expansion \eqref{eq:CrossChannel} can thus be rewritten as sums over $m,\mb=0,1,2,\dots$ for the family \eqref{eq:OHL}.

In the bulk, the kinematic regime that we are concerned with is that of high-energy scattering at fixed impact parameter. In the boundary theory this corresponds to studying 4-point correlators in the \emph{Regge limit}~\cite{Cornalba:2006xk, Cornalba:2006xm, Cornalba:2007zb, Cornalba:2007fs}, defined by an analytic continuation in one of the conformal cross-ratios (here chosen to be $z$) around the origin, followed by taking both $z$ and $\zb$ to $1$:
\begin{align} \label{eq:Rzzb}
	z\to z\, e^{-2\pi i}\ ,\qquad z,\zb\to1 \ .
\end{align}
This procedure is non-trivial since the analytic continuation crosses a branch cut of the hypergeometric functions in \eqref{eq:ConformalBlockDef} and so the limits $z,\zb\to1$ are taken on the second sheet of the correlator.
It is useful to parametrise the cross-ratios on the second sheet as \cite{Li:2017lmh}
\begin{align} \label{eq:SigEta}
	z = 1-\sigma \ ,\qquad \zb=1-\sigma\eta \ ,
\end{align}
in which case the Regge limit is obtained by taking $\sigma \to 0$ with $\eta$ held fixed.

Our goal is to study how the information about bulk high-energy scattering processes is encoded in the CFT. In particular, we focus on the cross-channel expansion \eqref{eq:CrossChannel} where the relevant data is contained in the anomalous dimensions $\Gamma_{m,\mb}$ and the OPE coefficients $C_{H\!L\mathcal{O}}$.
The exact comparison between the bulk and boundary is currently beyond our reach for the microstate geometries that we are interested in due to the fact that the relevant 4-point correlators are known only in terms of series. Since an analytic continuation is needed to go to the Regge regime of a correlator, we have to work perturbatively by expanding in the parameter \cite{Kulaxizi:2018dxo, Karlsson:2019qfi, Giusto:2020mup}
\begin{align} \label{eq:muDef}
    \mu \equiv \frac{24 h_H^{[0]}}{c} \ .
\end{align}
Such an expansion can be naturally interpreted in the bulk as an expansion in the mass of an AdS-Schwarzschild black hole, while on the CFT side it counts the number of copies of the stress tensor appearing in the Virasoro vacuum block \cite{Kulaxizi:2018dxo}.
Assuming that the anomalous dimensions and the three-point couplings admit an expansion in $\mu$ of the form%
\footnote{These anomalous dimensions should be thought of as `averages' over all operators in the spectrum that are degenerate at leading order in $1/N$. No lifting of this degeneracy \cite{Aprile:2017xsp} is studied here.}
\begin{subequations}
	 \label{eq:DoubleTraceData}
\begin{align}
    \Gamma_{m,\mb} &= \sum_{n=1}^{\infty} \, \G{n} \,  \mu^n 
    \label{eq:AnomDimExp}
    \ ,\\*
    C^{\,2}_{m,\mb} &\equiv C_{H\!L\mathcal{O}_{H\!L}} C_{\bar{\mathcal{O}}_{H\!L}H\!L} = \C{0}\left( 1 + \sum_{n=1}^{\infty}\, \C{n}\,  \mu^n  \right)
    \label{eq:OPECoefExp}
     \ ,
\end{align}
\end{subequations}
where we have suppressed the $m$ and $\mb$ dependence of the various series coefficients and have extracted the generalised free theory (GFT) OPE coefficients $\C{0}$, whose form can be found for instance in \cite{Heemskerk:2009pn}.

Inserting these series into the cross-channel decomposition \eqref{eq:CrossChannel} and expanding in $\mu$, 
one finds that in general the expansion coefficients $\G{n}$ and $\C{n}$ are encoded in the correlator in a complicated manner.
However, the analysis can be somewhat simplified by taking the Regge limit.
The cross-channel decomposition of an analytically continued correlator can be written to leading order in large $N$ as
\begin{align}\label{eq:CrossRegge}
	 \cC(z,\zb)
	&\xlongrightarrow[]{\circlearrowright} z^{-h^{[0]}_H-h_L}\, \bar{z}^{-\hb^{[0]}_H-\hb_L} \sum_{m,\mb=0}^{\infty} C^{\,2}_{m,\mb}\,e^{-\pi i \,\Gamma_{m,\mb}} g^{h_{H\!L},\hb_{H\!L}}_{h,\hb}(z,\zb)\,,
\end{align}
where we sum over all members of the family of exchanged HL multi-trace operators \eqref{eq:OHL}. In the above sum, the global conformal blocks implicitly depend on $\mu$ via the conformal dimensions $h$ and $\hb$ given in \eqref{eq:DTConDim}.
After taking $z, \zb\to 1$ a large number of terms contribute in the cross-channel, however, the dominant multi-trace operators are those with $m,\mb\gg1$. Correspondingly, we can treat $m$ and $\mb$ as continuous variables and approximate the sums by integrals. 
Then using \eqref{eq:DoubleTraceData}, one can write the explicit expansion of the correlator in the Regge limit, which we denote by $\cC_R$, as
\begin{align} \label{eq:CrossReggeMu}
	 \cC_{\rm R} &\approx \frac{1}{z^{h^{[0]}_H+h_L}\, \bar{z}^{\hb^{[0]}_H+\hb_L}} \int_0^{\infty}\!dm\!\int_0^{\infty}\!d\mb\, \C{0} \Bigg[ 1 + \mu\Big(\C{1} + \G{1}\Db\Big)+ \mu^2\bigg(\C{2}  + \frac{1}{2}\G{1}^2\Db^2\nonumber \\*
	 &\quad \qquad +\Big(\C{1}\G{1}+\G{2}\Big)\Db \bigg)+ \mu^3\bigg( \C{3}+ \frac{1}{6}\G{1}^3\Db^3 +  \frac{1}{2}\G{1}\Big( \C{1}\G{1}+2\G{2}\Big)\Db^2
		 \nonumber\\*
		 &\quad \qquad +\Big(\C{2}\G{1}+\C{1}\G{2} + \G{3}\Big)\Db \bigg)\Bigg]\,g^{h_{H\!L},\hb_{H\!L}}_{h,\hb}(z,\zb)\Bigg|_{\mu =0}
\end{align}
where $\Db\equiv \tfrac{1}{2}(\pd_m+\pd_{\mb})-\pi i$ and the the conformal block functions should be evaluated at $\mu =0$ only after the action of the derivatives in $\Db$.
Finally, when explicitly evaluating such integrals one can use the scaling limit $h_H\gg m,\mb\gg1$ in which the GFT OPE coefficients simplify to
\begin{align} \label{eq:C0HR}
	\C{0}(m,\mb) \approx \frac{m^{2 h_L-1} \mb^{2 \hb_L-1}}{\Gamma(2h_L)\,\Gamma(2\hb_L)} \ ,
\end{align}
and, since the hypergeometric functions are unity up to order $O(h_H^{-1})$ corrections, the global conformal blocks \eqref{eq:ConformalBlockDef} can be approximated by
\begin{align} \label{eq:GlobalHR}
	g^{h_{H\!L},\hb_{H\!L}}_{h,\hb}(z,\zb) \approx z^h\zb^{\hb} \,, 
\end{align}
with $h$ and $\hb$ being given by \eqref{eq:DTConDim}.

The link between the bulk phase shift \eqref{eq:BulkPS2} and the CFT data \eqref{eq:DTConDim} is made by using the impact parameter representation of 4-point correlation functions in the Regge limit \cite{Cornalba:2006xm,Kulaxizi:2018dxo,Karlsson:2019qfi}.
In what follows we limit our analysis to correlators involving only operators for which $h_L=\hb_L$ and $h_H=\hb_H$, as is the case for the explicit examples considered in section~\ref{sec:conical} and section~\ref{sec:twocharge}.
Starting from the cross-channel decomposition \eqref{eq:CrossChannel},
we want to write the correlator in the Regge limit as an expansion over impact parameter partial waves $\mathcal{I}_{m,\mb}$ as
\begin{align} \label{eq:ImpCross}
	 \cC_{\rm R}(z,\zb) \approx \sum_{m=0}^{\infty}\sum_{\mb=0}^{m} \mathcal{I}_{m,\mb} \, A(m,\mb) \ ,
\end{align}
where $A(m,\mb)$ is an arbitrary symmetric function of $m$ and $\mb$ such that $A|_{\mu=0}=1$.
In $d=2$, we define 
\begin{align} \label{eq:ImpPartial}
    \mathcal{I}_{m,\mb} \equiv 
   \frac{\C{0}(m,\mb)}{\abs{z}^{2(h^{[0]}_H+h_L)}} \left.g^{h_{H\!L},h_{H\!L}}_{h,\hb}(z,\zb)\right|_{\mu=0} + (z\leftrightarrow\zb)
    \approx \frac{(m\mb)^{2 h_L-1} }{\Gamma^2(2h_L)}\big( z^m\zb^{\mb} + z^{\mb}\zb^m\big) \ ,
\end{align}
where we have used \eqref{eq:C0HR} and \eqref{eq:GlobalHR} in the final right-hand side.
The impact parameter partial waves are defined to be the GFT ($\mu =0$) cross-channel partial waves of the operators \eqref{eq:OHL}. Hence by setting $\mu=0$ in \eqref{eq:ImpCross} we recover the disconnected part of the correlator in the Regge limit using
\begin{align} \label{eq:mu0ImpResum}
    \sum_{m=0}^{\infty}\sum_{\mb=0}^{m} \mathcal{I}_{m,\mb} \approx \abs{1-z}^{-4h_L} \ .
\end{align}
In \cite{Cornalba:2006xm} it was shown that $\mathcal{I}_{m,\mb}$ admits an impact parameter representation, which in $d=2$ and in the HHLL regime takes the form\footnote{Note that our conventions follow \cite{Giusto:2020mup} and differ from those of \cite{Cornalba:2006xm} and \cite{Karlsson:2019qfi}.}
\begin{align} \label{eq:ImpactRep}
    \mathcal{I}_{m,\mb} = \frac{2^{5-4h_L}\pi^2}{\Gamma^2(2h_L)}\int_{M^-}\frac{d^2p}{(2\pi)^2} (-p^2)^{2h_L-1}\,e^{-ip\cdot x} \,\abs{m-\mb}\,\delta^2 \ ,
\end{align}
where the integral is over the lower Milne wedge $M^- \equiv\{\left.\mathbf{p}\in\mathbb{R}^{1,1}\,\right|p^2\leq0,p_t\leq0\}$ and 
$\delta^2$ is defined as the combination of delta functions 
\begin{align} \label{eq:delta2}
	\delta^2 \equiv \delta(p\cdot\bar{e} +m+\mb)\,\delta\Big(\frac{p^2}{4}+m\mb\Big) \ ,
\end{align}
with $\bar{e}_{\mu} \equiv -\delta_{\mu}^{\,0}$. To show that the impact parameter representation \eqref{eq:ImpactRep} is equivalent to  \eqref{eq:ImpPartial}, we first evaluate the $p$ integrals using lightcone coordinates $p_{\pm} = p_t\pm p_y$ and $x^{\pm} = t\pm y$ yielding
\begin{align} \label{eq:MilneIint}
    \mathcal{I}_{m,\mb} = \frac{2^{2-4h_L}}{\Gamma^2(2h_L)}\int_{0}^{\infty}\!\!\!\int_{0}^{\infty} dp_+dp_-\, (p_+p_-)^{2h_L-1}\,e^{-\tfrac{1}{2}i(p_+x^-+\,p_-x^+)}\,\abs{m-\mb}\, \delta^2 \ ,
\end{align}
where the delta functions can be put into the form
\begin{align} \label{eq:deltafns}
	\delta^2 = \abs{m-\mb}^{-1}\Big(\delta\big(\tfrac{1}{2}\pp\!+m\big)\delta\big(\tfrac{1}{2}\pn\!+\mb\big) + \delta\big(\tfrac{1}{2}\pp\!+\mb\big)\delta\big(\tfrac{1}{2}\pn\!+m \big)\Big) \ .
\end{align}
By defining the coordinates on the plane as
\begin{align}\label{eq:PlaneCoordDef}
	z\equiv e^{ix^+}\,, \qquad \zb\equiv e^{ix^-}\,,
\end{align} 
it is simple to evaluate the integrals in \eqref{eq:MilneIint} to give exactly \eqref{eq:ImpPartial}.

Equally, we can insert the impact parameter representation of $\cI_{m,\mb}$ into \eqref{eq:ImpCross} and approximate the sums over $m$ and $\mb$ with integrals.
By performing first the integrals over $m$ and $\mb$, one obtains
\begin{align} \label{eq:mmbInt}
     \cC_{\rm R}(z,\zb) &\approx \int_0^{\infty}\!dm\!\int_0^{m}\!d\mb\ 
     \mathcal{I}_{m,\mb}\, A(m,\mb) \nonumber\\*
    &= \frac{2^{5-4h_L}\pi^2}{\Gamma(2h_L)^2} \int_{M^-}\!\frac{d^2p}{(2\pi)^2}\,(-p^2)^{2h_L-1}\,e^{-ip\cdot x}\, A(-\tfrac1{2}\bp) \,,
\end{align} 
with the identifications \eqref{eq:PlaneCoordDef} implied. 
We have used the notation $\bp \equiv (p_-, p_+)$ and the relations
\begin{align}
	\label{eq:Ppmtommb}
	\frac{\pp}{2} = -\mb\ \ ,\quad\frac{\pn}{2} = -m\,,
\end{align}
which originate from the delta functions in \eqref{eq:deltafns}.
The final line of \eqref{eq:mmbInt} can be inverted and if we set $A(m,\mb) = e^{i\delta(m,\mb)}$ then the phase can be determined in terms of the Fourier transform of a 4-point function in the Regge limit
\begin{align}
	\int\!d^2x\, \cC_{\rm R}(z,\zb)\,e^{ip\cdot x} = B_0(\mathbf{p})\,e^{i\delta(\mathbf{p})} \ ,
\end{align}
with  
\begin{align}
	B_0(\bp) = \frac{2^{3-4h_L}\,\pi^2}{\Gamma^2(2h_L)}\Theta(-p_t)\Theta(-p^2)\left(-p^2\right)^{2h_L-1}\ ,
\end{align}
denoting the Fourier transform of the disconnected part of the correlator.  
Following the proposal of \cite{Kulaxizi:2018dxo}, we identify $\delta(\bp)$ as the bulk phase shift of \eqref{eq:BulkPS} and interpret 
$\mathbf{p}$ as the conserved momentum along the null geodesics.
In particular, from \eqref{eq:Ppmtommb} it follows that
\begin{align} \label{eq:ptpyRels}
 \abs{p_t} = m+\mb\,,\qquad \ p_y = m-\mb \,,
\end{align}
and thus $\beta = -(m-\mb)/(m+\mb)$. This identification, valid in the Regge limit, directly links well-defined bulk quantities and the operators exchanged in 4-point correlators and will allow us to obtain concrete relations between the phase shift and CFT data \eqref{eq:DoubleTraceData}.
To obtain expressions linking the two sides, we insert \eqref{eq:ImpPartial} and $A(m,\mb) = e^{i\delta(m, \mb)}$ into \eqref{eq:ImpCross}, approximate the sums over $m$ and $\mb$ with integrals and use the assumed symmetry of the phase shift to get
\begin{align} \label{eq:CorrRexp}
     \cC_{\rm R}(z,\zb) \approx \int_0^{\infty}\!\!\!\int_0^{\infty}\!dmd\mb\,\frac{(m\mb)^{2h_L-1}}{\Gamma^2(2h_L)}\,z^m\zb^{\mb}\,e^{i\delta(m, \mb)} \ .
\end{align}
This relation incorporates the key idea of the eikonal regime; namely that by comparing with the GFT results \eqref{eq:C0HR} and \eqref{eq:GlobalHR} (valid in the Regge limit), one notes that the details of the interaction between the light probe and the heavy object are resummed into the phase. 
In the context of the proposal of \cite{Kulaxizi:2018dxo},  \eqref{eq:CorrRexp} directly relates the bulk phase shift $\delta(m,\mb)$ (with the relations \eqref{eq:ptpyRels} understood) to the CFT data \eqref{eq:DoubleTraceData}. Therefore, this relation allows us to reproduce the Regge limit of particular HHLL correlators and to extract cross-channel CFT data, all purely from the bulk phase shift of the probe propagating in the geometry dual to the heavy operator.

To see this explicitly, we assume that the phase shift admits the perturbative expansion%
\footnote{To be precise, the bulk phase shift that is related to the CFT data is the difference in the phase shift relative to the pure AdS result.
Therefore, the expansion of $\delta$ starts at linear order in $\mu$, since the contribution from the pure AdS is first subtracted.}
\begin{align}
	\label{eq:PhaShiExp}
	\delta(m, \mb)= \sum_{n=1}^{\infty} \delta_{(n)} \,\mu^n\,,
\end{align}
where from now on we suppress the $m$ and $\mb$ dependence of both the expansion coefficients $\delta_{(k)}$ and the exact phase shift $\delta$, and expand the right-hand side of \eqref{eq:CorrRexp} using
\begin{align} \label{eq:eexp}
	e^{i\delta} \approx 1 + i\D{1}\, \mu + \left(i\D{2} - \frac{1}{2}\D{1}^{2}\right)\mu^2 +\left(i\D{3} -\D{1}\D{2} -\frac{i}{6}\D{1}^3\right)\mu^3 \ .
\end{align}
Comparing the result term by term with the Regge limit cross-channel conformal block decomposition \eqref{eq:CrossReggeMu} one gets a series of differential relations between CFT data and the bulk phase shift. Since only elastic scattering is considered here, we take $\delta$ and all of its expansion coefficients to be real-valued. We similarly assume that the anomalous dimension $\Gamma_{m, \mb}$ and the square of the OPE coefficients $C^{\,2}_{m, \mb}$ take real values. Then the set of differential equations can be split by considering the real and imaginary parts separately, the first few of which yield previously known relations \cite{Cornalba:2006xm, Kulaxizi:2018dxo, Karlsson:2019qfi}%
\footnote{There are additional total derivative terms which we omit in these expressions.
Their contributions vanish in all examples we have considered and furthermore, the equations \eqref{eq:mu1relations} have been checked in the example of AdS-Schwarzschild and agree with the results obtained by independent lightcone limit methods \cite{Kulaxizi:2018dxo, Karlsson:2019qfi}.} 
\begin{subequations}
\label{eq:mu1relations}
\begin{align} 
\G{1} &= - \frac{\D{1}}{\pi}\ ,& \,\C{0}\C{1} &= \pdp \Big(\C{0}\G{1}\Big) \ ,\\*
\G{2} &= -\frac{\D{2}}{\pi} + \G{1}\,\pdp\G{1}\ ,& \C{0}\C{2} &= \pdp\!\left(\C{0} \C{1} \G{1} +\C{0}\G{2}\right) -\frac{1}{2} \pdp^2\!\left(\G{1}^{\,2}\,\C{0}\right) \ ,
\end{align}
\end{subequations}
where we have defined the differential operator $\pdp \equiv \partial_{(m +\mb)} = \frac12\left( \pd_m+\pd_{\mb}\right)$.
These relations quickly become cumbersome, particularly since they express CFT information in terms of both the phase shift expansion coefficients and lower order CFT data.
However, as might be expected from the point of view of holography, the bulk and boundary information can be separated and so it is possible to express the anomalous dimensions and OPE coefficients purely in terms of the bulk phase shift. 
The first few such relations read
\begin{subequations} \label{eq:GammaRels}
	\begin{align}
		\G{1} & = - \frac{\D{1}}\pi \,,\\*
		\G{2} &=  - \frac{\D{2}}{\pi} + \frac1{2!} \pdp\!\left(\frac{\D{1}^2}{\pi^2} \right)\,,\\*
		\G{3} &= - \frac{\D{3}}{\pi}  +  \pdp\!\left(\frac{\D{1}\, \D{2}}{\pi^2}\right) - \frac{1}{3!}\pdp^2\!\left(\frac{\D{1}^3}{\pi^3}\right)\ ,
	\end{align}
\end{subequations}
and 
\begin{subequations} \label{eq:CRels}
	\begin{align}
		\C{1} & = - \frac1{\C{0}}\pdp \left(\C{0} \frac{\D{1}}\pi\right) \,,\\*
		\C{2} &=  - \frac1{\C{0}} \pdp\!  \left[\C{0} \frac{\D{2}}{\pi} - \frac1{2!} \pdp\!\left(\C{0}\frac{\D{1}^2}{\pi^2} \right)\right]\,,\\*
		\C{3} &= - \frac1{\C{0}} \pdp\! \left[\C{0}\frac{\D{3}}{\pi}- \pdp\! \left(\C{0}\frac{\D{1}\, \D{2}}{\pi^2}\right) + \frac{1}{3!}\pdp^2\! \left(\C{0}\frac{\D{1}^3}{\pi^3}\right) \right] \ .
	\end{align}
\end{subequations}
A pattern soon emerges and so we conjecture that at an arbitrary order in $\mu$ these relations read 
\begin{subequations}
	\label{eq:GnCn}
	\begin{align}
		\label{eq:Gn}
		\G{n} &=  \sum_{\{k_t\}} \pd_+^{K-1} \!\left[\,\prod_{t = 1}^{\infty} \frac{1}{k_t!}\left(-\frac{\delta_{(t)}}{\pi}\right)^{\!\! k_t} \right]\,,\\*
		\label{eq:Cn}
		\C{n} &= \frac{1}{\C{0}}\sum_{\{k_t\}}  \pd_+^{K} \!\left[ \C{0} \prod_{t = 1}^{\infty} \frac{1}{k_t!}\left(-\frac{\delta_{(t)}}{\pi}\right)^{\!\! k_t} \right] ,
	\end{align} 
\end{subequations}
where the sums run over all configurations of $k_t \in\mathbb{N}^0$ such that
\begin{align}
    n = \sum_{t=1}^{\infty} t k_t\quad \text{and}\quad K = \sum_{t=1}^{\infty} k_t \ .
\end{align}
What this means is that in \eqref{eq:GnCn}, for each value of $n\in \mathbb{N}$ one sums over all integer partitions of $n$ with each term being divided by the degeneracy of the parts appearing in the partition.

These equations were derived using expressions that are valid only in $d=2$, however, it was argued in \cite{Karlsson:2019qfi} that the relations \eqref{eq:mu1relations} hold in any dimension and thus we can conjecture that the same should be true for \eqref{eq:GnCn}.
Interestingly, we observe that for light operators with $h_L = \hb_L= \tfrac1{2}$, the zeroth order OPE coefficient \eqref{eq:C0HR} are $\C{0}\approx1$ in the Regge limit and as a consequence, the CFT data associated with such probes satisfies a simple relationship at each order in $\mu$
\begin{align}
	\label{eq:RelFer}
	\C{n} = \pdp \G{n}\,.
\end{align}

Finally, let us note that one can resum the series coefficients appearing in \eqref{eq:GnCn} to obtain relations that are exact in $\mu$
\begin{align} \label{eq:GCexact}
	\Gamma_{m,\mb} = \sum_{n=1}^{\infty} \frac1{n!}\,\pd_+^{n-1}\!\bigg(\!\!-\frac{\delta}{\pi}\,\bigg)^{\!\!n} \,,\qquad 
	C^{\,2}_{m,\mb} = \frac1{\C{0}}\sum_{n=1}^{\infty} \frac1{n!}\, \pd_+^{n}\! \bigg[\C{0} \bigg(\!\!-\frac{\delta}{\pi}\,\bigg)^{\!\! n}\,\bigg]\,.
\end{align}
While we do not use these exact expressions in this work, we will discuss their potential in section~\ref{sec:disc}.

\section{Conical defect geometry}
\label{sec:conical}

In this section we apply the methods of section~\ref{sec:back} to the case of heavy operators dual to conical defect geometries.
We begin by very briefly reviewing the bulk computation of the phase shift for this example, followed by an analysis of 4-point correlation functions for two types of light probes and find perfect agreement between the CFT data and bulk phase shift.
Finally, we briefly comment on the possibility of the Regge limit distinguishing between pure and mixed states.

\subsection{Bulk description}

Perhaps the simplest non-trivial testing ground for the ideas presented in the previous section is the spacetime which is locally AdS${}_3\times S^3$, but has a conical singularity of order $k \in \mathbb{N}$ at the origin. 
The metric on this spacetime can be written as \cite{Balasubramanian:2000rt,Maldacena:2000dr}
\begin{align} \label{eq:ConicalDefect}
	ds^{2}_k = \frac{\sqrt{Q_1Q_5\,}\,}{r^2 + \tfrac{a^2}{k^2}}\,dr^2 - \frac{r^2 + \tfrac{a^2}{k^2}}{\sqrt{Q_1Q_5\,}\,}\, dt^2 + \frac{r^2}{\sqrt{Q_1Q_5\,}\,}\, dy^2 + \sqrt{Q_1Q_5}\,\, ds^{\,2}_{\!S^3} \ .
\end{align}
In the above, $Q_1$ and $Q_5$ denote the charges associated with the D1- and D5-branes from the string-theoretic set-up which are related to the radii of the AdS${}_3$ and the $S^1_y$ via the relation $\sqrt{Q_1Q_5\,}=R_{\mathrm{AdS}}^2=a \,R_y\,$, with $a$ being a real-valued parameter.
The above metric can be analytically extended to non-integer values of $k$ and to make contact with the literature we define
\begin{align} \label{eq:kContinued}
	\alpha \equiv \frac{1}{k} = \sqrt{1-\mu\,} \ ,
\end{align}
where $0\leq \alpha \leq 1$ and $\alpha =1$ corresponding to pure AdS$_3\times S^3$, a spacetime with no conical deficit.
The parameter $0\leq \mu\leq 1$ introduced in the above expression will act as the expansion parameter in the perturbative approach since $\mu = 0$ corresponds to an undeformed spacetime.

We now briefly review the results of the calculation performed in \cite{Giusto:2020mup} of the bulk phase shift in this spacetime with the (analytically continued version of the) metric \eqref{eq:ConicalDefect}.
We study a null geodesic beginning and ending on the boundary of AdS with no momentum in the $S^3$ directions, in which case the three-dimensional reduced metric is trivially obtained by dropping the $ds^{\,2}_{\!S^3}$ term in \eqref{eq:ConicalDefect}.
For such geodesics, the radial turning point is found to be 
\begin{align} \label{eq:Defectr0}
	r_0= \alpha\, a\,\bigg(\frac{1}{\beta^{2}}-1\bigg)^{-1/2} \,,
\end{align}
and the bulk phase shift  is given by 
\begin{align} \label{eq:DefectPS}
	\delta &= \pi R_y\,|p_t|\big(1-|\beta|\big)\,\big(\alpha^{-1}-1\big) \,,
\end{align}
where we have already subtracted the pure AdS${}_3$ result.
Using \eqref{eq:ptpyRels} we can write this phase shift in the form
\begin{align} \label{eq:DefectPSmmb}
	\delta = 2\pi\min(m,\mb)\,\big(\alpha^{-1}-1\big) \approx \pi\min(m,\mb)\bigg[ \mu + \frac{3}{4}\mu^2 + \frac{5}{8}\mu^3 \bigg] \ ,
\end{align}
where $2\min(m, \mb) = m+ \mb - |m-\mb|$ and in the last step we used \eqref{eq:kContinued} and expanded in $\mu$.

\subsection{CFT description}

Now we turn to the study of 4-point functions involving heavy operators  dual to the geometry in \eqref{eq:ConicalDefect}. 
The relevant heavy operators belong to the set of Ramond-Ramond ground states of the D1-D5 CFT, each having conformal dimensions $h_H=\hb_H=\tfrac{N}{4}$ and $SU(2)$ charges 
$j_H=\bar{j}_H = \tfrac{N}{2k}$, and hence the corresponding reduced conformal dimension \eqref{eq:ReducedDim} is given by 
\begin{align} \label{eq:ConicalDim}
	h_H^{[0]} = \frac{N}{4}\bigg(1-\frac{1}{k^2}\bigg) \ ,
\end{align}
where in the last expression, as for the rest of this section unless explicitly stated otherwise, we have used the analytic continuation \eqref{eq:kContinued}.

Four-point functions in the supergravity limit involving such heavy operators were first analysed in \cite{Galliani:2016cai}.
In the case when the light probe is a chiral primary operator that we denote $\OL=\Ofer$ with $h_L = \hb_L = 1/2$ and $j_L = \jb_L =1/2$, the correlator (in the NSNS sector) can be written as 
\begin{equation} \label{eq:DefectCfer}
	\cC^{\mathrm{fer}}(z,\zb) = \frac{\alpha}{|1-z|^2}\frac{1-|z|^2}{1-|z|^{2\alpha}} \,.
\end{equation}
From this result we can also obtain the expression for a correlator involving another light operator $\OL=\Obos$ -- contained in the same supermultiplet as $\Ofer$, with $h = \hb=1$ but zero R-symmetry charge -- via the superconformal Ward identity \cite{Bombini:2017sge}
\begin{align} \label{eq:DefectCbos}
    \cC^{\mathrm{bos}}(z, \zb) = \pd\pdb\,\Big[\cC^{\mathrm{fer}}(z, \zb)\Big] \ .
\end{align}

We begin by analysing the Regge limit of the correlator with the chiral primary light operator \eqref{eq:DefectCfer}.
We use the prescription \eqref{eq:Rzzb}: first analytically continuing $z$ around the origin (where the branch cut of $z^{\alpha}$ is crossed) and then using the parametrisation in \eqref{eq:SigEta} to take $\sigma\to0$ with $\eta$ fixed. This gives
\begin{align} \label{eq:DefectCferR}
	\cC^{\mathrm{fer}}_{\rm R} &= \frac{\alpha}{|1-z|^2}\frac{1-|z|^2}{1-|z|^{2\alpha}e^{-2\pi i\alpha}}\Bigg|_{z, \zb\to1} \nonumber \\*
	&\approx \frac1{\eta\sigma^2} + \frac{\pi i}{\eta(1+\eta)\sigma^3}\,\mu  - \frac{\pi^2}{\eta(1+\eta)^2\sigma^4}\,\mu^2 -\frac{\pi^3i}{\eta(1+\eta)^3\sigma^5} \,\mu^3 \ ,
\end{align}
where we have kept only the leading term in the small $\sigma$ expansion at each order in $\mu$, which at order $\mu^k$ behaves as $\sigma^{-k-2}$.

As discussed in section~\ref{sec:back}, it should be possible to reconstruct the Regge limit of this CFT correlator from the bulk phase shift using \eqref{eq:CorrRexp}.
Using the phase shift computed for the conical defect \eqref{eq:DefectPSmmb}, expanding the exponential function of \eqref{eq:CorrRexp} in $\mu$ and taking the Regge limit gives the following expression%
\footnote{In order to reduce clutter in our expressions, we use the notation $\cC^{\mathrm{fer}}_{\mathrm{AdS}}$ to mean the correlator predicted by the bulk phase shift using \eqref{eq:CorrRexp}, which is only valid in the Regge limit.}
\begin{align} 
    \cC_{\rm AdS}^{\rm fer} &\approx I_{0,0} + \pi i\mu \,I_{0,1} + \Big(\pi i\,I_{0,1} -\frac{\pi^2}{2}I_{0,2}\Big)\,\mu^2 + \bigg( \frac{5\pi i}{8}I_{0,1} -\frac{3\pi^2}{4}I_{0,2} - \frac{\pi^3 i}{6}I_{0,3} \bigg)\,\mu^3\nonumber \\*
    &\approx  \frac1{\eta\sigma^2} + \frac{\pi i}{\eta(1+\eta)\sigma^3}\,\mu - \frac{\pi^2}{\eta(1+\eta)^2\sigma^4}\,\mu^2 -\frac{\pi^3i}{\eta(1+\eta)^3\sigma^5}\,\mu^3\,,
    	\label{eq:DefectFerPSR}
\end{align}
where in the first line we have used the integral%
\footnote{For the derivation of this result see appendix A of \cite{Giusto:2020mup}.}
\begin{align}
    I_{a,b}(z,\zb) \equiv \!\int_0^{\infty}\!\!\!\!\int_0^{\infty}\!\!dmd\mb\, m^a\mb^b\big(z^m\zb^{\mb} + z^{\mb}\zb^m\big) =  \frac{\Gamma(a+b+2)}{(b+1)}\,\big[F(z,\zb) + F(\zb,z)\big] \, ,
\end{align}
with $F(z,\zb)\equiv \big(\!\!-\!\log{z}\big)^{-a-b-2}{}_2F_1\Big(b+1,a+b+2;b+2;-\frac{\log{\zb}}{\log z}\Big)$
and in the second line of \eqref{eq:DefectFerPSR} we have gone to the Regge limit by using the parametrisation \eqref{eq:SigEta}, taking the leading contribution in the small $\sigma$ expansion at each order in $\mu$ separately.

Comparing the two methods, we see that the phase shift calculation \eqref{eq:DefectFerPSR} precisely reproduces the Regge limit of the exact correlator \eqref{eq:DefectCferR}.
The matching of the two results can be checked to higher orders in $\mu$ and we do not expect the agreement to cease at an arbitrary order in the expansion.
Note that taking the $\sigma \rightarrow 0$ limit is crucial here since the key relation \eqref{eq:CorrRexp} is valid only in the Regge limit. 
However, in the case of this conical defect correlator with chiral primary light operator, the matching extends beyond the first order terms in small $\sigma$ as one finds additional agreement between the subleading terms at each order in $\mu$.
Neglecting the disconnected part of the correlator,%
\footnote{We note that the order $\mu^0$ term is the disconnected part of the correlators and as such does not contain any information about the interaction between the operators. 
This term is exact in $\sigma$: for four-point functions with light operators with $h_L=\hb_L$, it is given by $\eta^{-2h_L}\sigma^{-4h_L}$.}
the two methods begin to differ at subsubleading order in $\sigma$ with the leading differences being 
\begin{align} \label{eq:DefectferRdiff}
   \Delta \cC^{\mathrm{fer}}_{\mathrm{R}}\approx -\frac{3(1+\eta)^2+\pi i(1+\eta+\eta^2)}{12\eta(1+\eta)\sigma}\,\mu - \frac{\pi^2 }{12(1+\eta)^2\sigma^2}\,\mu^2 - \frac{\pi^3i}{12(1+\eta)^3\sigma^3}\,\mu^3 \,,
\end{align}
where we have used the notation $\Delta \cC^{\mathrm{fer}}_{\mathrm{R}} = \cC_{\rm R}^{\rm fer}-\cC_{\rm AdS}^{\rm fer}$.

One can repeat the above procedure for the conical defect correlator with the light operator $\Obos$. 
Analytically continuing the correlator obtained via \eqref{eq:DefectCbos} and taking the small $\sigma$ limit yields 
\begin{align} \label{eq:Defectbosms}
	\cC^{\mathrm{bos}}_{\rm R} \approx \frac1{\eta^2\sigma^4} + \frac{2\pi i(1+3\eta+\eta^2)}{\eta^2(1+\eta)^3\sigma^5}\,\mu - \frac{3\pi^2 (1+4\eta+\eta^2)}{\eta^2(1+\eta)^4\sigma^6}\,\mu^2 - \frac{4\pi^3 i(1+5\eta+\eta^2)}{\eta^2(1+\eta)^5\sigma^7}\,\mu^3 \ ,
\end{align}
where at each order in $\mu$ we have again kept only the leading contribution as $\sigma \to 0$, which now scales as $\sigma^{-k-4}$ at order $\mu^k$.
Obtaining the Regge limit from the bulk phase shift follows in the same way as in the case of the chiral primary light operator, however, since in this case $h_L=\hb_L=1$ the GFT OPE coefficients appearing in \eqref{eq:CorrRexp} need to be modified appropriately. This simply has the effect of replacing each integral in \eqref{eq:DefectFerPSR} by $\cI_{a,b}\to\cI_{a+1,b+1}$.
After taking the relevant limit, we find that the leading order again precisely matches the CFT result \eqref{eq:Defectbosms}.
However, unlike in the case of $\OL=\Ofer$, the discrepancy between the two approaches already sets in at the subleading term in $\sigma$ at each order in $\mu$, giving
\begin{align} \label{eq:DefectbosRdiff}
	\Delta \cC^{\mathrm{bos}}_{\mathrm{R}} \approx \frac{2\pi i(1+3\eta+\eta^2)}{\eta^2(1+\eta)^2\sigma^4}\,\mu - \frac{3\pi^2 (1+4\eta+\eta^2)}{\eta^2(1+\eta)^3\sigma^5}\,\mu^2 - \frac{4\pi^3 i(1+5\eta+\eta^2)}{\eta^2(1+\eta)^4\sigma^6}\,\mu^3\ .
\end{align}

Let us note here that one can check that at each order in $\mu$, the small $\eta$ limit of the expansions of the correlator expressions \eqref{eq:DefectCferR} and \eqref{eq:Defectbosms} agree with the results of the Virasoro vacuum block with $h_L=\hb_L=\tfrac1{2}$ and $h_L=\hb_L=1$ respectively.

As shown in section \ref{ssec:CFTPic}, the bulk phase shift can also be used to extract CFT data in the Regge limit. Using the expansion of the bulk phase shift \eqref{eq:DefectPSmmb} in the relations \eqref{eq:GammaRels} and \eqref{eq:CRels} gives the data
\begin{equation} \label{eq:DefectData}
\begin{aligned}
    \G{1} &\approx -\min(m,\mb)\,,\qquad& \G{2} &\approx -\frac{\min(m,\mb)}{4}\,,\qquad& \G{3}&\approx -\frac{\min(m,\mb)}{8}\,,\\
    \C{1} &\approx -\frac1{2}\,, & \C{2} &\approx -\frac1{8}\,, & \C{3} &\approx -\frac1{16}\,,
\end{aligned}
\end{equation}
where the OPE coefficients are calculated for the chiral primary light operator.
The exact anomalous dimensions are known from considerations of the bulk energy levels of the light probe in the conical defect geometry \cite{Kulaxizi:2018dxo}, which agree with our extracted data.

Finally, we comment about the analytic continuation \eqref{eq:kContinued} that we have used throughout the chapter. 
As noted previously, the conical defect geometry is dual to a pure state only when the parameter $k$ is a positive integer.%
\footnote{This can be seen, for example, by considering the D1-D5 CFT at the free orbifold point in which the heavy operator dual to the conical defect geometry with $k\in\mathbb{N}$ is related through a spectral flow transformation to $N/k$ copies of an order $k$ twist operator \cite{Lunin:2001jy}.}
Despite this, the correlator \eqref{eq:DefectCfer} after analytic continuation has the desired behaviour in the lightcone OPE limit ($\zb\to1$), reducing to the affine Virasoro vacuum block.
Hence, for non-integer values of $k$ one may be able interpret the geometry as being dual to an ensemble of heavy states, with the analytically continued values for the conformal dimensions and R-symmetry charges treated as averages over the ensemble.

One of the aims of this paper is to investigate whether it is possible use the Regge limit to distinguish between pure and mixed states. 
Previous results \cite{Kulaxizi:2018dxo, Karlsson:2019qfi, Giusto:2020mup} suggest that the analysis at first order in $\mu$ cannot separate mixed from pure states and what we find at higher orders in $\mu$ agrees with this.
On the flip side, this indicates the robustness of the proposal \eqref{eq:CorrRexp} in the sense that the bulk phase shift can reliably reconstruct the Regge limit of boundary correlators, even if the heavy operators are not pure states.

\section{Two-charge black hole microstate}
\label{sec:twocharge}

In this section we analyse high-energy scattering in a particular smooth microstate geometry of the two-charge D1-D5 black hole.
The geometry, first constructed in \cite{Kanitscheider:2007wq}, has a known dual description in terms of a pure heavy state in the D1-D5 CFT. 
Some 4-point correlation functions involving these heavy states are known exactly in terms of a double Fourier series \cite{Bombini:2017sge} and we use this result to analyse the Regge limit beyond the leading order in the deformation parameter $\mu$.
Finally, we compare and contrast our results to the expressions obtained in the conical defect geometry.

\subsection{Bulk description}

Let us consider the following three-dimensional line element 
\begin{align} \label{eq:3d100Metric}
	ds^{\,2} = \sqrt{Q_1Q_5\,} \,\frac{r^2+\frac{a^4}{a_0^2}}{(r^2 + a^2)^2 \,} \,dr^2- \frac{r^2+\frac{a^4}{a_0^2}}{\sqrt{Q_1Q_5\,}}\,dt^2 +\frac{r^2}{\sqrt{Q_1Q_5\,}}\, dy^2 \ .
\end{align}
This geometry is smooth and asymptotes to AdS${}_3$ with radius $R_{AdS}=(Q_1 Q_5)^{1/4}$ and has been extracted from a more complicated six-dimensional spacetime \cite{Bombini:2017sge} by considering the Kaluza-Klein ansatz form for the reduction on $S^3$ \cite{Giusto:2020mup}.
The six-dimensional geometry is sometimes referred to as the $(1,0,0)$ geometry, and with abuse of nomenclature we will use the same name for the dimensionally reduced spacetime.%
\footnote{The full six-dimensional spacetime from which we obtain \eqref{eq:3d100Metric} is the simplest of the $(k,0,0)$ class of microstate geometries considered in \cite{Bombini:2017sge}. 
While the full six-dimensional spacetime does not locally factorise into AdS$_3\times S^3$, one can consider a Kaluza-Klein ansatz for the reduction on $S^3$ and find that for $k=1$ the asymptotically AdS$_3$ part of the metric does not depend on the coordinates of the three-sphere. For that case, the resulting three-dimensional Einstein frame metric is \eqref{eq:3d100Metric}. For a more detailed analysis, see \cite{Giusto:2020mup}.}
Note that in addition to the fixed parameters $Q_1$, $Q_5$, and $a_0$, related through $\sqrt{Q_1 Q_5} = a_0 R_y\,$, the metric contains an additional (continuous) free parameter $a$ which determines the deviation from pure AdS$_3$.
In fact, we often use another parameter $b$, related to $a$ and $a_0$ through the constraint
\begin{align}
	a_0^2 = a^2 + \frac{b^2}{2}\,,
\end{align}
which ensures that the geometry is smooth.
It is easy to see that $a = a_0$ or equivalently $b=0$ corresponds to pure AdS, while if $a = 0$ (or $b^2 = 2a_0^2$) the metric \eqref{eq:3d100Metric} becomes that of a massless extremal BTZ black hole \cite{Banados:1992gq, Banados:1992wn}. 

Studying the null geodesics that start and end on the asymptotic AdS boundary, one finds that the phase shift is given by \cite{Giusto:2020mup}
\begin{align}
	\label{eq:100PS}
	\delta &=\pi R_y\,|p_t|\,\Bigg( \sqrt{\frac{2a_0^2-b^2\beta^2}{2a_0^2-b^2} \,}-1\,\Bigg)\,,
\end{align}
where we have already subtracted the contribution from pure AdS. 
Interestingly, the radial turning point in this geometry is given by 
\begin{align}
	r_0 = \frac{a^2}{a_0} \big(\beta^{-2}-1\big)^{-\frac{1}{2}} \,,
\end{align}
and has the same form as in the conical defect case \eqref{eq:Defectr0}. Specifically, they have a factorised form between a parameter determining the deviation from pure AdS (either $\alpha$ or $a^2/\textsf{}a_0^2$) and a shared function of $\beta$ which denotes the dependence on the choice of geodesic. 
Despite this, there is no similarity in the expressions of the phase shift: in the case of the conical defect we see a factorisation of the phase shift in $\beta$ and $\alpha$ dependence \eqref{eq:DefectPS}, whereas here there is no such separation in \eqref{eq:100PS} for the $(1,0,0)$ geometry.

The expansion parameter $\mu$ is in this case defined through the relation
\begin{align} \label{eq:bmuRel}
	\sqrt{1-\mu\,} = \frac{a^2}{a_0^2} = 1- \frac{b^2}{2a_0^2}\ ,
\end{align}
which allows us to expand the phase shift \eqref{eq:100PS} in powers of $\mu$, with the first few orders reading
\begin{small}
\begin{align}
	\frac{\delta}{\pi} &\approx \frac{m\mb}{m+\mb}\,\mu 
	 + \left( \frac{3m \mb}{4(m+\mb)} - \frac{m^2\mb^2}{2(m+\mb)^3}\right)\!\mu^2 
	 + \left(\frac{5 m \mb}{8 (m+\mb)} - \frac{3m^2\mb^2}{4(m+\mb)^3} + \frac{m^3\mb^3}{2(m+\mb)^5}\right)\!\mu^3 \,,
\end{align}
\end{small}
where we used \eqref{eq:ptpyRels} to relate the momenta to $m$ and $\mb$.

\subsection{CFT description}

The six-dimensional geometry that can be reduced to \eqref{eq:3d100Metric} is dual to a pure heavy state in the D1-D5 CFT: an RR sector ground state with conformal dimensions $h_H=\hb_H=\tfrac{N}{4}$ and R-symmetry charges $j_H=\bar{j}_H=\tfrac1{2}(N-N_b)$, with the corresponding reduced conformal dimension thus being
\begin{align} \label{eq:100Dim}
	h_H^{[0]} = \frac{N_b}{2}\bigg(1-\frac{N_b}{2N}\bigg) \,.
\end{align}
The additional free parameter $N_b$ appearing in the expression of the charges is related to the free parameters of the bulk description through \cite{Bena:2017xbt}
\begin{align}
	\label{eq:100Ntob}
	\frac{N_b}{N} = \frac{b^2}{2a_0^2} \qquad \Longrightarrow \qquad 1- \frac{N_b}{N} = \frac{a^2}{a_0^2}\ ,
\end{align}
which, when inserted into \eqref{eq:100Dim} and using the definition \eqref{eq:muDef}, justifies the relation \eqref{eq:bmuRel} on the gravity side.\footnote{The heavy states dual to the geometry \eqref{eq:3d100Metric} are related by a spectral flow transformation to a state in the NS-NS sector, composed of $N-N_b$ copies of the vacuum state and a coherent superposition of $N_b$ (with scaling \eqref{eq:HeavyScaling}) copies of an anti-chiral primary of dimension $h_{\rm NS}=\hb_{\rm NS}=\tfrac1{2}$.
This heuristically justifies the relation \eqref{eq:100Ntob}, since from this CFT point of view $N_b$ can be considered a measure of the deviation from the vacuum state, which is holographically dual to global AdS${}_3\times S^3$.}
We note that in order for the operator to be dual to a semiclassical background representing a fully backreacted geometry and not just a small perturbation on top of global AdS, we require the `heavy' scaling
\begin{align} \label{eq:HeavyScaling}
    N_b\sim O(N)\ \text{ as }\ N\to\infty\ ,
\end{align}
where the ratio $N_b/N$ is fixed and left undetermined. 
Our perturbative analysis in $\mu$ is in essence an expansion in this free parameter.%
\footnote{Even though the scaling \eqref{eq:HeavyScaling} is essential in this holographic approach, there is evidence that the limit $N_b \to 0$ is smooth and yields valid results for correlators with only light operator insertions \cite{Giusto:2018ovt, Giusto:2019pxc, Giusto:2020neo, CGHR}.}

The 4-point HHLL correlation functions where the heavy operators are dual to the geometry \eqref{eq:3d100Metric} were studied in \cite{Galliani:2017jlg} to first order in the ratio $\tfrac{b^2}{2a_0^2}$ while in \cite{Bombini:2017sge} an exact expression in terms of a double Fourier series was found.
In the case of the light operator being the chiral primary $\OL=\Ofer$ with $h_L = \hb_L = \frac12$, the correlator is given by%
\begin{align}
	\label{eq:100CorFerzzb}
	\cC^{\rm fer}(z, \zb) = \frac{a}{a_0} \sum_{k=1}^{\infty}\,\sum_{l\in \mathbb{Z}}\frac{1}{\sqrt{1+ \frac{b^2}{2a^2}\, \frac{l^2}{\left(|l|+2k\right)^2}}}\,\left(\frac{z}{\zb}\right)^{\frac{l}{2}} \, \left(z \zb\right)^{-\frac{a}{2 a_0}\sqrt{\left(|l|+2k\right)^2 + \frac{b^2}{2a^2} l^2}}\,.
\end{align}
We can use \eqref{eq:bmuRel} and expand this result as a series in $\mu$
\begin{align}
	\label{eq:100CorFerSer}
	\cC^{\rm fer} (z, \zb) = \sum_{n=1}^{\infty} \mu^{n} \, \cC_n^{\rm fer}(z, \zb)\,,
\end{align}
where now the expressions for $\cC^{\rm fer}_n(z, \zb)$ can be written in a closed form in terms of Bloch-Wigner-Ramakrishnan polylogarithm functions \cite{zagier1991polylogarithms}.
The properties of these correlators are analysed in more detail in \cite{CGHR} but a quick summary, together with their explicit form, can be found in the Appendix~\ref{app:corr}.
Here we are interested purely in their behaviour in the Regge limit.
Using the explicit forms of the correlators \eqref{eq:AExplicitCorr} one finds that in the Regge limit the correlator with $\OL=\Ofer$ is given to leading order in $\sigma$ by
\begin{align}
	\label{eq:CferCFT}
	\cC^{\mathrm{fer}}_{\rm R} &\approx \frac{\pi i\big(1-\eta^2+2\eta\log\eta\big)}{(1-\eta)^3\eta\sigma^3}\,\mu
	 - \frac{\pi^2\big(1+9\eta-9\eta^2-\eta^3 + 6\eta(1+\eta)\log\eta\big)}{(1-\eta)^5\eta\sigma^4}\,\mu^2 \nonumber\\*
	&\quad -\frac{\pi^3i\big(1+28\eta-28\eta^3-\eta^4 + 12\eta(1+3\eta+\eta^2)\log\eta\big)}{(1-\eta)^7\eta\sigma^5}\,\mu^3 \ .
\end{align}

On the other hand, using the bulk phase shift \eqref{eq:100PS} in \eqref{eq:CorrRexp} and expanding in $\mu$ yields (in the Regge limit) 
\begin{align} \label{eq:CferRPS}
	\cC^{\rm fer}_{\rm AdS} &\approx \Bigg[\pi i\,\cIt_{1,1}\, \mu  + \left( \frac{3\pi i}{4}\,\cIt_{1,1} - \frac{2\pi i}{4}\,\cIt_{2,3} - \frac{\pi^2}{2}\,\cIt_{2,2}\right)\,\mu^2
	  \nonumber\\*
	  &	\quad +  \left(\frac{5\pi i}{8}\,\cIt_{1,1} - \frac{3\pi i}{4}\,\cIt_{2,3} +\frac{\pi i}{2}\,\cIt_{3,5}- \frac{\pi^3i}{6}\,\cIt_{3,3} - \frac{3\pi^2}{4}\,\cIt_{2,2} + \frac{\pi^2}{2}\,\cIt_{3,4}\right)\,\mu^3\Bigg]\Bigg|_{z, \zb\to1}\,,
\end{align}
where we have made use of the integral
\begin{align}
	\label{eq:ITilde}
	\cIt_{a,b} \equiv \int_0^{\infty}\!\!\!\int_0^{\infty}\!dmd\mb\, z^m\zb^{\mb}\frac{(m\mb)^a}{(m+\mb)^b} =  \frac{\Gamma^2(a+1)\,\Gamma(2a+2-b)}{\Gamma(2a+2)\,\big(\!\!-\!\log\zb\big)^{a+1-b}}\ \widetilde{F}(z,\zb) \ ,
\end{align}
with $\widetilde{F}(z,\zb) = \big(\!-\log z\big)^{-a-1} {}_2F_1\Big(a+1,b\,;2a+2\,;1-\frac{\log\zb}{\log z}\Big)$.
After using the parametrisation \eqref{eq:SigEta}, the correlator predicted from the bulk phase shift agrees completely with \eqref{eq:CferCFT} to leading order in $\sigma$, at each order in $\mu$.
In fact, the matching between these two results persists until order $\sigma^{-1}$ for all terms in the $\mu$ expansion, with the leading difference being 
\begin{align}
		\Delta \cC^{\mathrm{fer}}_{\mathrm{R}} \approx -\frac{(3+i \pi)(1+\eta)}{12 \eta \sigma }\, \mu - \frac{(3 - i \pi)(1+\eta)}{48\eta \sigma}\, \mu^2 - \frac{(45 - 15 \pi i + 2 \pi^3 i)(1+\eta)}{1440 \eta \sigma}\, \mu^3\,,
\end{align}
where we have again neglected the difference in the disconnected term. 
This breakdown of validity of the Regge limit approximation is very different in nature to that occurring in the conical defect correlator \eqref{eq:DefectferRdiff} case, especially as it seems that in this case the matching becomes increasingly better as we increase the order in $\mu$. 
In principle the reconstruction of the correlator from the phase shift is expected to be valid only in the Regge limit, or in other words, at leading order in $\sigma$. It would be interesting to explain this enhancement of the reconstruction and see whether it might allow us to probe interactions beyond the Regge regime. However, we must note that this difference in matching between the $(1,0,0)$ fuzzball geometry and the conical defect cannot be seen if the light operator is $\OL=\Obos$ and the corresponding 4-point correlation function is related to \eqref{eq:100CorFerzzb} via the Ward identity \eqref{eq:DefectCbos}.
In that case -- analysed in more detail in Appendix~\ref{app:corr} -- 
the difference between the Regge limit of the correlator and the reconstruction from the bulk phase shift starts as expected at the subleading contribution in $\sigma$ at each order in $\mu$, just as in the example of the conical defect \eqref{eq:DefectbosRdiff}. 

Finally, we perform a check on the validity of the relations \eqref{eq:Gn}. 
In \cite{Bombini:2017sge} the exact energy levels of the composite object formed of the light probe and the heavy operator dual to the geometry \eqref{eq:3d100Metric} were calculated. 
Since the anomalous dimensions can be thought of as the bulk binding energies of such bound states, we can find the anomalous dimensions as%
\footnote{Note that in \cite{Bombini:2017sge} the expression is given in terms of $n = \min(m, \mb) + 1$ and $l = m-\mb$.}
\begin{small}
\begin{align}
	\label{eq.BindingEnergy}
	&\Gamma_{m,\mb} =\omega_{m, \mb} - \omega_{m, \mb}\big|_{b\to0} =   \frac{a}{a_0}\sqrt{ (m+\mb+2)^2 + (m-\mb)^2\frac{b^2}{2a^2}\,}- (m+\mb+2) \nonumber\\
	& \approx - \frac{m\mb}{m+\mb}\mu - \frac{m\mb(m^2 + 4m\mb+\mb^2)}{4(m+\mb)^3}\mu^2 - \frac{m \mb (m^4 + 6 m^3 \mb+14 m^2 \mb^2 + 6 m \mb^3 +\mb^4) }{8(m+\mb)^5}\,\mu^3  \ ,
\end{align}
\end{small}%
where in the second line we have expanded the result in $\mu$ and taken the large $m$ and $\mb$ limit relevant for the Regge regime. 
One can then compare the results at each order in $\mu$ to those obtained from the phase shift \eqref{eq:100PS} and \eqref{eq:Gn} and find perfect agreement to arbitrary orders in $\mu$.
For the case of $\OL=\Ofer$ we can then use the simple relation \eqref{eq:RelFer} to get the following OPE coefficients
\begin{align}
	\C{1} &\approx - \frac{m^2+\mb^2}{2(m +\mb)^2}\ , \qquad\qquad   \C{2}  \approx - \frac{m^4 + 6 m^3 \mb - 2 m^2 \mb^2 + 6 m \mb^3 + \mb^4}{8 (m + \mb)^4}\nonumber \\
	\C{3}&\approx  - \frac{m^6 + 8 m^5 \mb + 23 m^4 \mb^2 - 8 m^3 \mb^3 +23 m^2 \mb^4 + 8 m \mb^5 + \mb^6}{16 (m +\mb)^6}\ .
\end{align}

\section{Discussion}
\label{sec:disc}


In summary, we performed the explicit Regge limit analysis for two different examples: in section \ref{sec:conical} for the case of heavy operators dual to AdS${}_3$ conical defect geometries, analytically continued to real-valued deficits, and in section~\ref{sec:twocharge} for the $(1,0,0)$ family of microstate geometries.
We find that it is possible to reproduce the leading Regge limit of HHLL correlators and to extract CFT data of certain cross-channel HL multi-trace operators at arbitrary orders in the $\mu$ expansion
solely from the information contained in the bulk phase shift -- computed via a geodesic approximation of a light probe in the geometry dual to the heavy state. 
Furthermore, the perturbative analysis that we employ -- which effectively includes a checking of the crossing relations in the Regge limit -- appear to be unable to distinguish between mixed and pure states.

For the conical defect, the extension to higher orders in $\mu$ is straightforward since both the correlator and the phase shift were already computed in closed forms \cite{Galliani:2016cai, Giusto:2020mup}.
Our perturbative analysis confirmed that the Regge limit of the correlator is consistent at all orders in $\mu$, even if the defect parameter is analytically continued to values associated with mixed states.
In the case of the $(1,0,0)$ microstates, while the phase shift was already known to all orders in $\mu$ from \cite{Giusto:2020mup}, HHLL correlators involving this heavy operator were known in closed form only at first order. In this paper we use the closed form expressions at higher orders in $\mu$ recently found in \cite{CGHR} and, as in the conical defect case, our analysis of these $(1,0,0)$ correlators found no inconsistencies in the Regge limit.

This suggests that the Regge limit analysis, particularly the relation between the bulk and CFT data \eqref{eq:CorrRexp}, seems to be robust and consistent to arbitrary order in the perturbation series in $\mu$. One of our main objectives has been to prepare the ground for future work in the black hole regime where $\mu$ is large and so this is encouraging. In this regime it was found that the phase shift of a light probe in the BTZ black hole geometry (which has $\mu>1$) is ill-defined for all values of the impact parameter \cite{Kulaxizi:2018dxo}, owing to the fact that all in-falling null geodesics that begin on the asymptotic boundary fall into the black hole.
In contrast, asymptotically AdS black holes in higher dimensions do not suffer from such issues \cite{Kulaxizi:2018dxo} and one generically finds regions of both elastic and inelastic scattering \cite{Parnachev:2020zbr}.

It would be interesting to extend the Regge limit analysis to more complicated microstate geometries, particularly those that resemble black holes to an arbitrarily close degree.
Specifically, a line of future work is to study microstates of the 3-charge D1-D5-P black hole which has a finite horizon radius and thus represents a more realistic (albeit still extremal) black hole.
This analysis was initiated in \cite{Giusto:2020mup}, where the so-called $(1,0,1)$ microstate was studied at first order in $\mu$. 
One of the key observations is that, in contrast to the BTZ black hole, for horizonless fuzzball geometries considered thus far there seems to always exist a radial turning point for in-falling null geodesics and hence a well-defined phase shift.
It would be interesting to see whether one could study classical black holes as a particular limit of microstate geometries and analyse the transition between the purely elastic scattering found in fuzzballs and purely inelastic capture seen for BTZ black holes. We expect that in order to fully analyse this black hole regime it will be necessary to work exactly in $\mu$. 
A first attempt in that direction has been made in \eqref{eq:GCexact} where we relate the bulk phase shift on one hand and anomalous dimensions and OPE coefficients on the other, none of which are expanded in $\mu$. There is a subtlety in the above statements: in general the 3-charge microstate geometries are not dual to quasi-primaries and so our analysis seemingly cannot be applied to these cases. However, it was found in \cite{Giusto:2020mup} that at linear order in $\mu$ and in the heavy scaling regime \eqref{eq:HeavyScaling}, the naive application of the Regge analysis still holds. One can then hope that this also continues to hold true for higher orders. The analysis and implications of these ideas are left for future work.
Finally, we note that the derivation of these equations implicitly assumes that all quantities involved are real, or in other words that the scattering in the bulk is elastic. As discussed above, this is not always the case due to the inelastic part of the scattering process (such as capture of the light probe and radiative dissipation) typically being encoded in an imaginary component of the phase shift -- seen for example in \cite{Parnachev:2020zbr}. It would be interesting to extend \eqref{eq:GCexact} in this spirit and explore the implications.

\section*{Acknowledgements}

We would like to thank Rodolfo Russo and Stefano Giusto for collaboration in the early stages of this work and for helpful comments on the draft.
We would also like to thank the organisers of the conference \emph{Black-Hole Microstructure II}, where some of these ideas were presented.
The work of NC is supported by the ERC Grant 787320 - QBH Structure while the work of MH is supported by the Science and Technology Facilities Council (STFC) Consolidated Grant ST/P000754/1 {\it String theory, gauge theory \& duality}.
%
\appendix 
\section{Detailed analysis of the $(1,0,0)$ correlator}
\label{app:corr}

In this appendix, we gather some of the more technical details that are omitted from the main text. In particular, we explicitly state the first few terms of the series expansion \eqref{eq:100CorFerSer} and introduce the Bloch-Wigner-Ramakrishnan polylogarithm functions that represent the natural language in which to express these results. 
We also present the Regge limit analysis of the 4-point correlation function in the $(1,0,0)$ fuzzball geometry with a light probe of conformal dimensions $h_L = \hb_L =1$, which was omitted in section~\ref{sec:twocharge}.


\subsection{Series expansion of the full correlator}

We begin by analysing the HHLL correlator involving a chiral primary light operator and the heavy state dual to the $(1,0,0)$ microstate geometry. 
The exact expression was derived in \cite{Bombini:2017sge} by extracting the 2-point function $\langle\OL(z)\OLb(0)\rangle_H$ from the normalisable part of the regular solution to the wave equation of a minimally coupled scalar field in the geometry dual to $\OH$  (see \cite{Galliani:2016cai, Galliani:2017jlg} for more details).
In terms of the dimensionless coordinates on the cylinder $\tau={t}/{R_y}$ and $\sigma={y}/{R_y}$, the correlator is given by
\begin{equation} \label{eq:Ab1sumln}
	\mathcal{C}^\mathrm{fer}(\tau,\sigma)=
	\,\frac{a}{a_0}\sum_{k=1}^\infty\, \sum_{l\in\mathbb{Z}} e^{il\sigma} \frac{\exp\left[-i\frac{a}{a_0}\sqrt{(|l| + 2 k)^2+ \frac{b^2  l^2}{2 a^2}}\,\tau\right]}{\sqrt{1+\frac{b^2}{2 a^2}\frac{l^2}{(|l|+2k)^2}} } \,,
\end{equation}
where in order to make a connection to the expression \eqref{eq:100CorFerzzb} we can transform the coordinates to the Minkowski plane as 
\begin{equation}
	z = e^{i(\tau+\sigma)}\quad,\quad \zb= e^{i(\tau-\sigma)} \ .
\end{equation}

As discussed in more detail in \cite{CGHR}, the correlator \eqref{eq:Ab1sumln} admits a natural expansion in the ratio $b^2/{2a_0^2}$, in which case the series coefficients can be written in a closed form.
Here on the other hand, we find it more useful to expand in the parameter $\mu$ where
\begin{align}
	\label{eq:Amub}
	\frac{b^2}{2a_0^2} = 1 - \sqrt{1-\mu}\,,
\end{align}
so that the correlator takes on the form\footnote{Because we expand in $\mu$ rather than $b^2/{2a_0^2}$, the expressions in \eqref{eq:AExplicitCorr} slightly differ from those presented in \cite{CGHR}. 
However, the two sets of results are simply linear combinations of each other, determined by the relation \eqref{eq:Amub}.}
\begin{align}
	\label{eq:A100CorFerSer}
	\cC^{\rm fer} (z, \zb) = \sum_{n=1}^{\infty} \mu^{n} \, {\cC}_n^{\rm fer}(z, \zb)\,,
\end{align}
with the series coefficients being 
\begin{small}
	\begin{subequations}
		\label{eq:AExplicitCorr}
		\begin{align}
			\cC^{\rm fer}_0 & = \frac{1}{|1-z|^2}\,,\\
			\cC^{\rm fer}_1& =- \frac{1}{2|1-z|^2}-  \frac{|z|^2(z + \zbar - 2 z \zbar)}{2(z-\zbar)^2|1-z|^2} \log|z|^2  - \frac{|z|^2 }{(z-\zbar)^2}\log|1-z|^2 - \frac{2i |z|^2 (z+ \zbar) }{(z-\zbar)^3}P_2(z, \zbar)\,,\\
			\cC^{\mathrm{fer}}_2 &= \frac{1}{8}\bigg[ 
			- \frac{1}{|1-z|^2}+  \frac{\abs{z}^2(z+\zb-2\abs{z}^2)}{(z-\zb)^2\abs{1-z}^2}\,\log\abs{z}^2 + \frac{\abs{z}^2(z+\zb)}{(z-\zb)^2\abs{1-z}^2}\,\big(\log\abs{z}^2\big)^2 \nonumber \\*
			&  \qquad 
			 +\frac{2\abs{z}^2}{(z-\zb)^2}\log\abs{1-z}^2  + \frac{28i\abs{z}^2(z+\zb)}{(z-\zb)^3}\,P_2(z, \zbar)
			 - \frac{48\abs{z}^2(z^2+4\abs{z}^2+\zb^2)}{(z-\zb)^4}\,P_3(z, \zbar)\nonumber \\*
			&\qquad  +\frac{24i\abs{z}^2(z+\zb)(z^2+10\abs{z}^2+\zb^2)}{(z-\zb)^5}\,P_4(z, \zbar)\bigg] \,,\\*
			\cC^{\mathrm{fer}}_3 &= \frac{1}{16}\bigg[\!
			- \frac{1}{|1-z|^2}+ \frac{\abs{z}^2(z+\zb-2\abs{z}^2)}{(z-\zb)^2\abs{1-z}^2}\,\log\abs{z}^2  - \frac{\abs{z}^2(z+\zb)}{2(z-\zb)^2\abs{1-z}^2}\big(\log\abs{z}^2\big)^2
			 \nonumber\\*
			 &\qquad - \frac{2\abs{z}^2(z^2+4\abs{z}^2+\zb^2)\big(\log\abs{z}^2\big)^2}{3(z-\zb)^4}\,\bigg(\frac{z+\zb-2\abs{z}^2}{2\abs{1-z}^2}\log\abs{z}^2+  \log\abs{1-z}^2 \bigg) \nonumber\\
			& \qquad  +\frac{2\abs{z}^2}{(z-\zb)^2}\log\abs{1-z}^2 + \frac{4i\abs{z}^2(z+\zb)}{(z-\zb)^3}\,P_2(z, \zbar)+ \frac{80\abs{z}^2(z^2+4\abs{z}^2+\zb^2)}{(z-\zb)^4}\,P_3(z, \zbar)  \nonumber\\
			&\qquad -\frac{4i\abs{z}^2(z+\zb)(z^2+10\abs{z}^2+\zb^2)}{(z-\zb)^5}\,\Big(60P_4(z, \zbar)+\big(\log\abs{z}^2\big)^2\,P_2(z, \zbar)\Big) \nonumber\\
			&\qquad + \frac{4\abs{z}^2(z^4+26z^3\zb+66z^2\zb^2+26z\zb^3+\zb^4)}{(z-\zb)^6}\,\Big(60P_5(z, \zbar)+\big(\log\abs{z}^2\big)^2\,P_3(z, \zbar)\Big) \nonumber\\*
			&\qquad  -\frac{4i\abs{z}^2(z+\zb)(z^4+56z^3\zb+246z^2\zb^2+56z\zb^3+\zb^4)}{3(z-\zb)^7}\Big(60P_6(z, \zbar)+\big(\log\abs{z}^2\big)^2 \,P_4(z, \zbar)\Big)\! \bigg]\,.
		\end{align}
	\end{subequations}
\end{small}%
In the above expressions we have used the so called \emph{Bloch-Wigner-Ramakrishnan polylogarithm} functions\footnote{Here we only present the core properties of the generalised polylogarithm functions that are used in the main part of the text. A more in-depth presentation can be seen in \cite{CGHR} and in some of the original texts covering this topic \cite{Lewin:100000,Ramakrishnan:1986, zagier1990bloch, zagier1991polylogarithms, zagier:2006dilog}.}
$P_n(z, \zb)$, which are a higher-order generalisation of the usual Bloch-Wigner dilogarithm function $D(z, \zb)$, given in our notation by $P_2(z, \zb)$. 
These functions are defined as 
\begin{align}
	\label{eq:APmdef}
	P_n(z, \zb) &\equiv  \mathfrak{R}_n \left( \sum_{k = 0}^{n-1} \frac{ 2^k\,B_k}{k!}\, \left( \log|z|\right)^k\, \plog{n-k}{z}\right)\,,
\end{align}
with $\mathfrak{R}_n$ denoting the real or imaginary part for $n$ odd or even respectively, while the coefficients $B_j$ are the Bernoulli numbers.\footnote{The first few non-zero Bernoulli numbers are $B_0 = 1$, $B_1 = -\frac12$, $B_2 = \frac16$, $B_4 = - \frac1{30}$, $B_6 = \frac{1}{42}$, $B_8 =  -\frac{1}{30}$. All Bernoulli numbers with an odd valued index vanish, except for $B_1$.}
Using the definition \eqref{eq:APmdef}, one finds the following explicit expressions for the first few functions to be
\begin{small}
\begin{subequations}
	\label{eq:APdef}
	\begin{align}
		P_2(z, \zb) & =  \frac{1}{2i}\left[ \plog{2}{z} - \plog{2}{\zb} + \log|z|\, \log\left( \frac{1-z}{1-\zb}\right) \right]\,,\\
		P_3(z, \zb) &= \frac12 \left[ \plog{3}{z} + \plog{3}{\zb} - \log|z|\, \left( \plog{2}{z} + \plog{2}{\zb}\right) - \frac{2}{3} \left( \log|z|\right)^2\, \log|1-z| \right]\,,\\
		P_4(z, \zb) &= \frac{1}{2i}\left[ \plog{4}{z} - \plog{4}{\zb} - \log|z|\,\left( \plog{3}{z} - \plog{3}{\zb}\right) + \frac{1}{3}\left(\log|z|\right)^2 \left( \plog{2}{z} - \plog{2}{\zb} \right) \right]\,,\\
		P_5(z, \zb) &=  \frac12\left[ \plog{5}{z} + \plog{5}{\zb} - \log|z|\left( \plog{4}{z} + \plog{4}{\zb}\right)+ \frac13 \left(\log|z|\right)^2 \,\left(\plog{3}{z} + \plog{3}{\zb}\right)\right. \nonumber\\*
		& \quad \quad\left. + \frac{2}{45}\left(\log|z|\right)^4 \log|1-z| \right]
		\,,\\
		P_6(z, \zb) &=  \frac{1}{2i}\left[  \plog{6}{z} -\plog{6}{\zb} - \log|z| \left(\plog{5}{z} - \plog{5}{\zb} \right) + \frac13 \left(\log|z|\right)^2 \left(\plog{4}{z}-\plog{4}{\zb}\right)
		\right.\nonumber\\*
		&\left. \qquad \quad  - \frac{1}{45} \left(\log|z|\right)^4 \left(\plog{2}{z} -\plog{2}{\zb}\right)  \right]
		\,.
	\end{align}
\end{subequations}
\end{small}%
It is also important to note that these particular combinations of polylogarithm functions have some favourable symmetry properties. 
At a general order $n$ the functions satisfy an inversion relation
\begin{align}
	P_n(z, \zb) = (-1)^{n-1} P_n\left( \frac{1}{z}, \frac{1}{\zb}\right) \,,
\end{align}
while the standard Bloch-Wigner dilogarithm ($n=2$) satisfies additional symmetry properties  
\begin{align}
	\label{eq:Dsym}
	P_2(z, \zb) &= 	-P_2\left(\frac1{z}, \frac1{\zb}\right)  = - P_2(1-z, 1-\zb)= P_2\left( \frac{z-1}{z}, \frac{\zb-1}{\zb}\right)\nonumber \\
	&= P_2\left( \frac{1}{1-z}, \frac{1}{1-\zb}\right)= - P_2\left( \frac{z}{z-1}, \frac{\zb}{\zb-1}\right)\,.
\end{align}
Furthermore, if $\zb= z^*$ ({\it{i.e.}} if $P_n$ can be treated as a function of a single complex variable), then the above functions are real-analytic on the complex plane bar the points $z=0$ and $z=1$ where they are only continuous.
%


\subsection{Regge limit of the $(1,0,0)$ correlator with $\OL=\Obos$}

Here we analyse the Regge limit of the 4-point function of the correlator involving the heavy state dual to the $(1,0,0)$ geometry presented in section~\ref{sec:twocharge} and light operator $\OL=\Obos$ with $h_L = \hb_L = 1$.
This correlator is again related to the that involving the chiral primary light operator \eqref{eq:100CorFerzzb} through a superconformal Ward identity $\cC^{\mathrm{bos}} = \pd\pdb\left[\cC^{\mathrm{fer}}\right]$ \cite{Bombini:2017sge}.
However, if we expand this correlator in a series as
\begin{align}
	\label{eq:ABosSerExp}
	\cC^{\rm bos}(z, \zb) = \sum_{n=0}^{\infty} \mu^n \, \cC^{\rm bos}_{n}(z, \zb)\,,
\end{align}
then we can apply the Ward identity at each order in the series expansion directly 
\begin{align}
	\label{eq:Ward100}
	\cC^{\mathrm{bos}}_n(z, \zb) = \pd\pdb\left[\cC^{\mathrm{fer}}_n(z, \zb)\right] \,,
\end{align}
and use the closed form expressions in \eqref{eq:AExplicitCorr}.

We can now repeat the Regge analysis for this 4-point function. 
Using the above method of obtaining the $\cC^{\mathrm{bos}}_n$, we find that analytically continuing the obtained results and taking $z, \zb\to 1$ gives 
\begin{small}
    \begin{align}
		\label{eq:100BosCorr}
		\cC^{\rm bos}_{\rm R} &\approx \frac{1}{\eta^2\sigma^4} + \frac{2 \pi i }{\eta^2 (1-\eta)^5 \sigma^5}\,\left(1 - 8 \eta + 8 \eta^3 - \eta^4 - 12 \eta^2 \log\eta\right)\, \mu \nonumber\\*
		&  - \frac{3\pi^2}{\eta^2 (1-\eta)^7 \sigma^6}(1-15 \eta - 80\eta^2 + 80 \eta^3 + 15 \eta^4 - \eta^5 -60 \eta^2 (1+ \eta)\log \eta)\, \mu^2 
		 \, 
		 \nonumber\\*
		 &  -\frac{4 i \pi^3}{\eta^2(1-\eta)^9\sigma^7}\left(1 - 24 \eta - 375 \eta^2+ 375 \eta^4 + 24 \eta^5 - \eta^6 - 60 \eta^2 (3 + 8 \eta + 3\eta^3) \log \eta\right)\mu^3.
    \end{align}
\end{small}%
On the other hand we can use the bulk phase shift \eqref{eq:100PS} and its expansion in $\mu$ to reconstruct the Regge limit of the CFT correlator using \eqref{eq:CorrRexp}. 
We find that the analysis is the same as before only that now, due to the conformal dimensions of $\OL=\Obos$ being $h_L = \hb_L = 1$, we need to replace $\cIt_{a,b}\to\cIt_{a+1,b+1}$ in \eqref{eq:CferRPS} which explicitly yields 
\begin{align}
	\label{eq:ABosPS}
		\cC^{\rm bos}_{\rm AdS} &\approx \bigg[\pi i\,\cIt_{2,2}\, \mu  + \left( \frac{3\pi i}{4}\,\cIt_{2,2} - \frac{2\pi i}{4}\,\cIt_{3,4} - \frac{\pi^2}{2}\,\cIt_{3,3}\right)\,\mu^2
	\nonumber\\*
	&	\quad +  \left(\frac{5\pi i}{8}\,\cIt_{2,2} - \frac{3\pi i}{4}\,\cIt_{3,4} +\frac{\pi i}{2}\,\cIt_{4,6}- \frac{\pi^3i}{6}\,\cIt_{4,4} - \frac{3\pi^2}{4}\,\cIt_{3,3} + \frac{\pi^2}{2}\,\cIt_{4,5}\right)\,\mu^3\bigg]\bigg|_{z, \zb\to 1}\,.
\end{align}
Explicitly evaluating the integrals \eqref{eq:ITilde} and executing the limit completely reproduces the leading $\sigma$ behaviour of \eqref{eq:100BosCorr} at each order in $\mu$.
However, for this correlator -- unlike the case of the $\OL=\Ofer$ light operator and the $(1,0,0)$ heavy state -- the matching between the bulk reconstruction and the CFT Regge limit analysis ceases already at the subleading order in $\sigma$ at each order in $\mu$.
In fact for the first few orders that we have analysed, we find that 
\begin{align}
	\cC^{\rm bos}_{\rm R} \approx \big(1 + \sigma (1+ \eta)\big) \cC^{\rm bos}_{\rm AdS}\ ,
\end{align}
where at each order in $\mu$ the difference starts at the subsubleading contribution in $\sigma$.
The same behaviour can also be seen in the case where the heavy operator is dual to the conical defect (compare \eqref{eq:Defectbosms} and \eqref{eq:DefectbosRdiff}).

\bibliographystyle{JHEP.bst}
\bibliography{RLv2.bib}

\end{document}